\documentclass[aps, prx, twocolumn, superscriptaddress, longbibliography]{revtex4-2}

\usepackage{url}
\usepackage{braket}
\usepackage{graphicx}
\usepackage{multirow}
\usepackage{amsthm}
\usepackage{soul}
\usepackage{verbatim}
\usepackage{bm}
\usepackage{outlines}
\usepackage[exponent-product=\cdot]{siunitx}
\usepackage[space]{grffile}
\usepackage[dvipsnames]{xcolor}
\usepackage{xr}
\usepackage{color}
\usepackage{tikz}
\usetikzlibrary{calc, positioning}
\usepackage{mathtools}
\usepackage{algorithm}
\usepackage{algorithmic}
\usepackage{amsmath}
\usepackage{amssymb}
\usepackage{comment}
\usepackage[english]{babel}
\usepackage{subfigure}
\usepackage{booktabs}

\newcommand\myshade{85}
\colorlet{mylinkcolor}{violet}
\colorlet{mycitecolor}{YellowOrange}
\colorlet{myurlcolor}{Aquamarine}
\usepackage{hyperref}
\hypersetup{
	linkcolor  = mylinkcolor!\myshade!black,
	citecolor  = mycitecolor!\myshade!black,
	urlcolor   = myurlcolor!\myshade!black,
	colorlinks = true,
}
\usepackage[capitalize]{cleveref}

\newcommand{\GHz}{\mathrm{GHz}}
\newcommand{\MHz}{\mathrm{MHz}}

\newcommand{\ns}{\mathrm{ns}}
\newcommand{\us}{\mu\mathrm{s}}

\newcommand{\quotes}[1]{``#1''}
\newcommand{\abs}[1]{\left|{#1}\right|}
\newcommand{\brkt}[1]{\left(#1\right)}
\newcommand{\bbrkt}[1]{\left[#1\right]}

\newcommand{\QEC}{QEC}

\newcommand{\logerrrate}{\varepsilon_{\mathrm{L}}}

\newcommand{\leakrate}{L_{1}}
\newcommand{\seeprate}{L_{2}}
\newcommand{\tcycle}{t_{\text{c}}}
\newcommand{\Tone}{T_{1}}
\newcommand{\Toneup}{T_{1}^{\uparrow}}
\newcommand{\Tphi}{T_{\phi}}
\newcommand{\tdephsweet}{T_{\phi, \text{max}}}
\newcommand{\tdephpark}{T_{\phi, \text{park}}}
\newcommand{\tdephint}{T_{\phi, \text{int}}}
\newcommand{\tint}{t_{\mathrm{int}}}
\newcommand{\tphasecorr}{t_{\mathrm{pc}}}
\newcommand{\trise}{t_{\mathrm{rise}}}

\newcommand{\lcphase}[1][]{\ifthenelse{\equal{#1}{}}{\phi^{\mathrm{L}}}{\phi^{\mathrm{L}}_{\mathrm{#1}}}}
\newcommand{\tpiLRU}{t_{\pi\text{-LRU}}}
\newcommand{\tresLRU}{t_\text{\resLRU}}
\newcommand{\freq}{\omega}
\newcommand{\barecoupling}{g}
\newcommand{\tcz}{t_{\CZ}}

\newcommand{\cycle}[1][]{\ifthenelse{\equal{#1}{}}{n}{n_{\mathrm{#1}}}}
\newcommand{\ztype}{Z}
\newcommand{\xtype}{X}
\newcommand{\ytype}{Y}
\newcommand{\idtype}{I}
\newcommand{\datatype}{D}

\newcommand{\MWPM}{MWPM}
\newcommand{\UB}{UB}

\newcommand{\meas}[1][]{\ifthenelse{\equal{#1}{}}{m}{m_{\mathrm{#1}}}}
\newcommand{\tmeas}{t_{\mathrm{\meas}}}
\newcommand{\tgate}{t_{\mathrm{gate}}}

\newcommand{\flux}{\mathrm{flux}}
\newcommand{\static}{\mathrm{stat}}

\newcommand{\anharm}{\alpha}
\newcommand{\CZ}{\mathrm{CZ}}

\newcommand{\leakcondphase}[1][]{\ifthenelse{\equal{#1}{}}{\phi^{\leaksub}}{\phi^{\leaksub}_{\mathrm{#1}}}}

\newcommand{\compsub}{\mathcal{C}}
\newcommand{\leaksub}{\mathcal{L}}

\newcommand{\Ttwo}{T_{2}}

\DeclareMathOperator{\tr}{Tr}
\newcommand{\norm}[1]{\left|\left|{#1}\right|\right|}

\DeclareMathOperator{\linspan}{span}

\newcommand{\addition}[1]{{\color{black}{#1}}}

\newcommand{\lifetime}[1][]{\ifthenelse{\equal{#1}{}}{l}{l_{#1}}}
\newcommand{\avlifetime}[1][]{\ifthenelse{\equal{#1}{}}{\bar{l}}{\bar{l}_{#1}}}
\newcommand{\avglife}{l^\mathcal{L}_\mathrm{avg}}
\newcommand{\GammaCL}{\Gamma_{\compsub\rightarrow \leaksub}}
\newcommand{\GammaLC}{\Gamma_{\leaksub\rightarrow \compsub}}

\newcommand{\avgphoton}{\bar{n}}
\newcommand{\diag}{\mathrm{D}}
\newcommand{\offdiag}{\mathrm{OD}}
\newcommand{\dressed}{D}
\newcommand{\ddressed}{DD}
\newcommand{\thermal}{\sigma_\mathrm{th}}
\newcommand{\leakpop}{p^{\ket{2}}}
\newcommand{\onepop}{p^{\ket{1}}}
\newcommand{\zeropop}{p^{\ket{0}}}
\newcommand{\leakredrate}{R}
\newcommand{\leakrateLRU}{\leakrate^\mathrm{LRU}}
\newcommand{\readoutp}[2]{p_M(#1|#2)}

\newcommand{\SW}{\text{SWT}}
\newcommand{\deltaq}{\delta^q}
\newcommand{\deltar}{\delta^r}
\newcommand{\ar}{a}
\newcommand{\bq}{b}
\newcommand{\tildedeltaq}{\tilde{\delta}^q}

\newcommand{\pulse}{\Omega}

\newcommand{\steady}{{\,\mathrm{ss}}}
\newcommand{\resLRU}{res-LRU}
\newcommand{\piLRU}{$\pi$-LRU}
\newcommand{\swapLRU}{swap-LRU}
\newcommand{\drivesub}{\mathcal{S}}
\newcommand{\avgprob}{\bar{p}^{\mathcal{L}}}
\newcommand{\logicalfid}{\mathcal{F}_\mathrm{L}}
\newcommand{\distance}{d}
\newcommand{\Tslot}{T_\mathrm{slot}}
\newcommand{\tpulse}{t_\mathrm{p}}

\begin{document}
	
	\title{A hardware-efficient leakage-reduction scheme for quantum error correction with superconducting transmon qubits}
	
	\newcommand{\QuTech}{\affiliation{QuTech, Delft University of Technology, P.O.~Box 5046, 2600 GA Delft, The Netherlands}}
	\newcommand{\Kavli}{\affiliation{Kavli Institute of Nanoscience, Delft University of Technology, P.O.~Box 5046, 2600 GA Delft, The Netherlands}}
	\newcommand{\JARA}{\affiliation{JARA Institute for Quantum Information, Forschungszentrum Juelich, D-52425 Juelich, Germany}}
	\newcommand{\TNO}{\affiliation{Netherlands Organisation for Applied Scientiﬁc Research (TNO), P.O.~Box 96864, 2509 JG The Hague, The Netherlands}}
	\newcommand{\lorentz}{\affiliation{Instituut-Lorentz, Universiteit Leiden, P.O.~Box 9506, 2300 RA Leiden, The Netherlands}}
	
	\author{F.~Battistel}\thanks{battistel.fra@protonmail.com}\QuTech
	\author{B.~M.~Varbanov}\QuTech
	\author{B.~M.~Terhal}\QuTech\JARA
	
	\date{\today}
	
	\begin{abstract}
		
		Leakage outside of the qubit computational subspace poses a threatening challenge to quantum error correction~(QEC).
		We propose a scheme using two leakage-reduction units~(LRUs) that mitigate these issues for a transmon-based surface code, without requiring an overhead in terms of hardware or QEC-cycle time as in previous proposals.
		For data qubits we consider a microwave drive to transfer leakage to the readout resonator, where it quickly decays, ensuring that this negligibly disturbs the computational states for realistic system parameters.
		For ancilla qubits we apply a $\ket{1}\leftrightarrow\ket{2}$ $\pi$~pulse conditioned on the measurement outcome.
		Using density-matrix simulations of the distance-3 surface code we show that the average leakage lifetime is reduced to almost~$1$~QEC~cycle, even when the LRUs are implemented with limited fidelity.
		Furthermore, we show that this leads to a significant reduction of the logical error rate.
		This LRU~scheme opens the prospect for near-term scalable QEC~demonstrations.
		
	\end{abstract}
	
	\maketitle
	
	
	Quantum computing has recently reached the milestone of quantum supremacy~\cite{Arute19} thanks to a series of improvements in qubit count~\cite{Jurcevic20,Egan20}, gate fidelities~\cite{Rol16,Chen16b,Barends14,Sheldon16b,Hong20,Rol19a,Negirneac20,Yan18,Foxen20,Kjaergaard20,Sung20,Harty14} and measurement fidelities~\cite{Jeffrey14,Bultink16,Heinsoo18}.
	The next major milestones include showing a quantum advantage~\cite{Bravyi18,Zhong20,Babbush20,Arute20} and demonstrating quantum error correction~(QEC)~\cite{Kelly15,Riste15,Takita16,Negnevitsky18,Bultink19,Andersen19,Andersen20,Marques21,Egan20,Chen21}.
	Errors accumulate over time in a quantum computer, leading to an entropy increase which severely hinders the accuracy of its output.
	Thus QEC~is necessary to correct errors and remove entropy from the computing system.
	If the overall physical error rate is below a certain noise threshold for a given QEC-code family, the logical error rate decreases exponentially with the code distance~$\distance$ at the price of a~$\mathrm{poly}(\distance)$ overhead, thus allowing to extend the computational time.
	Recently, small-size instances of error-detecting~\cite{Andersen20,Marques21} and error-correcting~\cite{Egan20} codes have been experimentally realized.
	To further demonstrate fault tolerance it is crucial to scale up these systems and show that larger distance codes consistently lead to lower logical error rates than smaller distance codes~\cite{Chen21}.
	
	Leakage outside of the computational subspace~\cite{Strauch03,DiCarlo09,Martinis14,Rol19a,Negirneac20,Foxen20,Hong20,Tripathi19,Babu20,Werninghaus20}, present in leading quantum-computing platforms such as superconducting qubits and trapped ions, poses a particularly threatening challenge to fault tolerance~\cite{Aliferis07,Fowler13,Ghosh13_B,Ghosh15,Kelly15,Suchara15,Brown18,Brown19,Brown19_B, Varbanov20,Brown20,Mcewen21}.
	Leakage can increase entropy by making measurement outcomes no longer point to the underlying errors and can effectively reduce the code distance~\cite{Varbanov20}.
	Furthermore, leakage can last for many QEC~cycles~\cite{Ghosh13_B}, making operations on a leaked qubit fail and possibly spread correlated errors through the code~\cite{Fowler13,Varbanov20,Chen21}.
	In particular, leakage falls outside the stabilizer formalism of~QEC as it cannot be decomposed in terms of Pauli errors.
	Stabilizer codes~\cite{Fowler12,Obrien17} and their decoders are thus typically ill-suited to deal with leakage, leading to a significant increase of the logical error rate~\cite{Suchara15,Brown19_B,Varbanov20}.
	\addition{If the average leakage lifetime~$\avglife$, that is, the average number of QEC~cycles that a qubit stays leaked (after leaking in the first place), fulfills} $\avglife=\mathcal{O}(1)$~QEC~cycles and~$\avglife\ll d$, then for low-enough error rates a threshold is likely to exist~\cite{Fowler13} as leakage would have a relatively local effect in space and time.
	Due to a finite energy-relaxation time, leakage does indeed last for~$\avglife=\mathcal{O}(1)$~QEC~cycles.
	However, in practice it is important how large~$\avglife$ is, since if it is low the noise threshold is expected to be higher.
	Shortening the relaxation time to reduce~$\avglife$ is not effective as this increases the physical error rate as well.
	
	A leakage-reduction unit~(LRU)~\cite{Aliferis07,Fowler13,Suchara15,Ghosh15,Brown20,Hayes19,Langrock20,Mcewen21} is an operation introducing a seepage mechanism besides that of the relaxation channel.
	A LRU converts leakage into regular (Pauli) errors and shortens the average leakage lifetime, ideally to 1~QEC~cycle.
	As discussed above, this is expected to lead to a higher noise threshold, but not as high as for the case without leakage, since the leakage rate effectively adds to the regular error rate thanks to the LRU.
	As an alternative to the use of LRUs, post-selection based on leakage detection has been adopted~\cite{Varbanov20} as a near-term method to reduce the logical error rate.
	While leakage detection could also be used to apply LRUs in a targeted way, post-selection is not scalable.
	By shortening the lifetime to~$\avglife=\mathcal{O}(1)\ll d$, the use of LRUs is instead a scalable approach.
	
	In its imperfect experimental implementation a LRU can either introduce extra Pauli errors or mistakenly induce leakage on a non-leaked qubit.
	Furthermore, in the context of the surface code the LRUs investigated so far~\cite{Suchara15,Ghosh15,Brown20} introduce an overhead in terms of hardware and QEC-cycle time.
	Specifically, these LRUs are variants of the \swapLRU, in which the qubits are swapped at the end of each QEC~cycle, taking alternatively the role of data and ancilla qubits.
	In this way every qubit is measured every 2~QEC~cycles.
	The core of the \swapLRU~is the fact that the measured qubits are reset to the computational subspace after the measurement.
	This can be accomplished by a scheme which unconditionally maps~$\ket{1}$ and~$\ket{2}$ (and possibly~$\ket{3}$~\cite{Mcewen21}) to~$\ket{0}$~\cite{Magnard18,Zeytinoglu15,Egger18}, or conditionally using real-time feedback~\cite{Riste12b,Andersen19}.
	Under the standard assumption that the SWAP~gates swap the states of two qubits only if none of them is leaked (which does not necessarily hold in experiment~\cite{Mcewen21}), $\avglife$~is ideally shortened to 2~QEC~cycles.
	On the downside, for the pipelined surface-code scheme in~\cite{Versluis17}, the pipeline is disrupted as qubits cannot be swapped until the measurement and reset operations are completed, leading overall to an increase up to~$50\%$ of the QEC-cycle time depending on the reset time.
	The extra $\CZ$~gates, needed to implement the SWAPs, can cause additional errors or leakage as the~$\CZ$ is the major source of leakage in transmons~\cite{Strauch03,DiCarlo09,Martinis14,Rol19a,Negirneac20,Foxen20,Hong20,Tripathi19}.
	Moreover, in the surface code an extra row of qubits is needed to perform all the SWAPs~\cite{Ghosh15}, which is a non-negligible overhead in the near term.
	All these issues increase the physical error rate by a considerable amount, thus requiring to increase the system size to compensate for that (assuming that the error rates are still below threshold).
	
	In this work we propose two separate LRUs for data and ancilla qubits which use resources already available on chip, namely the readout resonator for data qubits~(\resLRU) and a $\ket{1}\leftrightarrow\ket{2}$ $\pi$~pulse conditioned on the measurement outcome for ancilla qubits~(\piLRU).
	In particular, the use of the \resLRU~avoids the necessity to swap data and ancilla qubits to be able to reset the data qubits.
	The \resLRU~is a modification of the two-drive scheme in~\cite{Magnard18,Zeytinoglu15,Egger18} to a single drive to deplete only the population in~$\ket{2}$ but not~$\ket{1}$, making it a LRU rather than a reset scheme.
	We additionally show that this negligibly affects the coherence within the computational subspace in an experimentally accessible regime, with a low probability of mistakenly inducing leakage as long as the thermal population in the readout resonator is relatively small.
	This allows \addition{us to unconditionally use} \resLRU~in the surface code in every QEC~cycle.
	In the pipelined scheme~\cite{Versluis17} the \resLRU~easily fits within the time in which the data qubits are idling while the ancilla qubits are finishing to be measured.
	As the \piLRU~can be executed as a short pulse at the end of the measurement time with real-time feedback, our LRU scheme overall does not require any QEC-cycle time overhead.
	Using density-matrix simulations~\cite{quantumsim_website,Obrien17,Varbanov20} of the distance-3 surface code~(Surface-17), we show that the average leakage lifetime is reduced to almost 1~QEC~cycle when \resLRU~and \piLRU~with realistic performance are employed.
	Furthermore, compared to the case without LRUs, the logical error rate is reduced by up to~$30\%$.
	The proposed \resLRU~and \piLRU~can be straightforwardly adapted to QEC-code schemes other than~\cite{Versluis17} and the \resLRU~is potentially applicable to superconducting qubits with higher anharmonicity than transmons.
	The demonstrated reduction serves as evidence of scalability for our LRU~scheme, even though we cannot estimate a noise threshold as we have simulated only one size of the surface code.
	To explore larger codes it is necessary to use less computationally expensive simulations~\cite{Fowler13,Kelly15,Suchara15} that use a simplified version of our error model at the cost of losing some information contained in the density matrix.
	Furthermore, to optimize the noise threshold the LRUs can be supplied with a leakage-aware decoder~\cite{Fowler13,Kelly15,Suchara15,Stace10,Nagayama17,Auger17} that uses measurement information about leakage to better correct leakage-induced correlated errors.

	\section{Readout-resonator LRU}
	\label{sec:readres}
	
	The readout resonator has been used~\cite{Magnard18,Zeytinoglu15,Egger18} to reset a transmon qubit to the~$\ket{0}$ state, depleting the populations in~$\ket{1}$ and~$\ket{2}$.
	Targeting the $\ket{20}\leftrightarrow\ket{01}$ transition, with the notation $\ket{\mathrm{transmon},\mathrm{resonator}}$, those populations are swapped onto the readout resonator, where they quickly decay due to the strong coupling to the transmission-line environment.
	Ref.~\cite{Magnard18} uses two drives simultaneously while Refs.~\cite{Zeytinoglu15,Egger18} use these drives in a three-step process.
	Here we adapt these techniques to use a single drive in a single step to deplete the population in~$\ket{2}$ only.
	
	A LRU is defined~\cite{Aliferis07} as an operation such that 1)~the incoming leakage population is reduced after the application of the LRU, 2)~the induced leakage when applied to a non-leaked state is ideally~0.
	We thus ensure below that not only leakage is reduced but also that the effect that the drive has on a non-leaked transmon is as small as possible.
	
	\subsection{Transmon-resonator Hamiltonian}
	\label{sub:analytical_results_main}
	
	\begin{figure}
		\centering
		\includegraphics[width=\columnwidth]{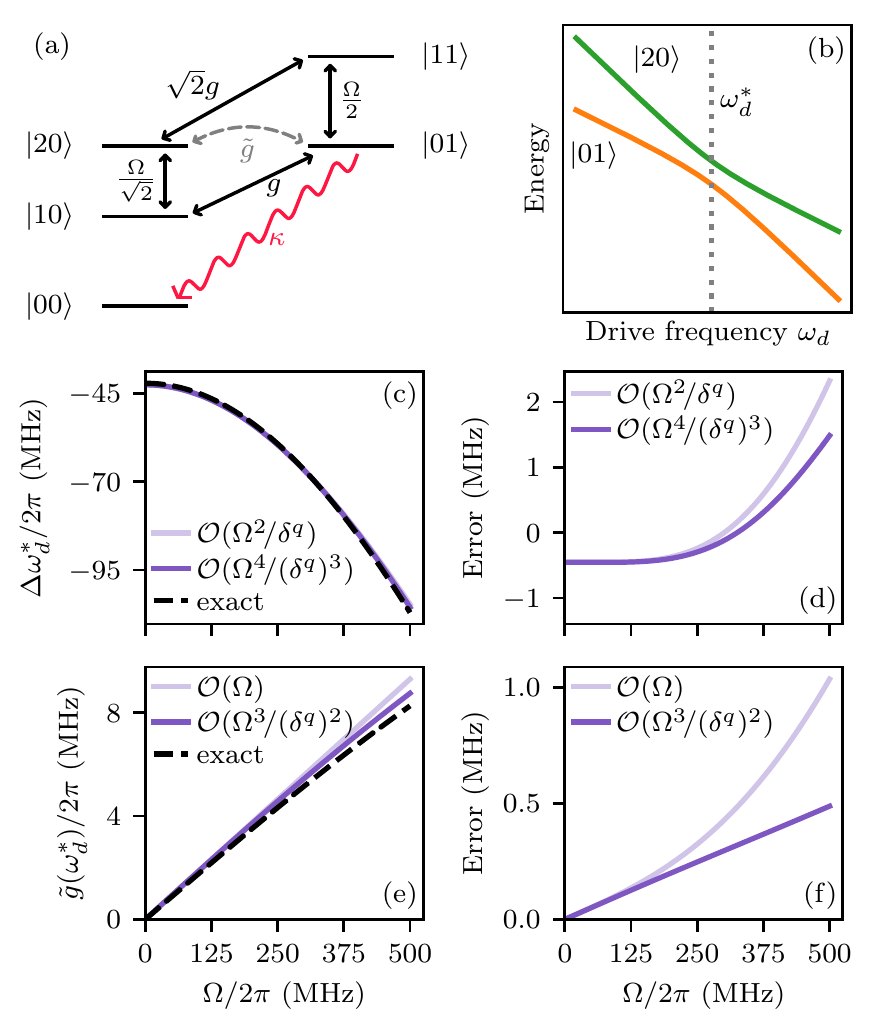}
		\caption{\label{fig1:readout_res_analytical}
			Concept of the readout-resonator LRU.
			(a)~The state~$\ket{20}$ (with the notation~$\ket{\mathrm{transmon,resonator}}$) is connected to~$\ket{01}$ by two main paths via either~$\ket{11}$ or~$\ket{10}$, due to the capacitive coupling~$g$ or the transmon-drive amplitude~$\Omega$, respectively.
			This generates an effective coupling~$\tilde{g}$ which can be used to swap~$\ket{20}\leftrightarrow\ket{01}$.
			The latter quickly decays to~$\ket{00}$ due to the typically high coupling~$\kappa$ of the readout resonator to the transmission-line environment, overall removing leakage from a leaked transmon.
			(b)~In the rotating frame of the drive, $\ket{20}$ and~$\ket{01}$ show an avoided crossing as a function of the drive frequency~$\omega_d$, centered at~$\omega_d^*$.
			The effective coupling~$\tilde{g}(\omega_d^*)$ is equal to half the energy separation at that point.
			(c),(e)~$\Delta\omega_d^*\coloneqq\omega_d^*-(2\omega_q+\anharm-\omega_r)$ and~$\tilde{g}(\omega_d^*)$ are respectively evaluated either exactly by full numerical diagonalization of~$H$ in~\cref{eq:H_full_main}, or by approximate analytical formulas (see~\cref{sub:analytical_results_main} and~\cref{sec:approx_hamil}) for the parameters in~\cref{tab:target_parameters}.
			\addition{The absolute errors with respect to the exact curves are shown in~(d),(f) respectively.}
		}
	\end{figure}
	
	We consider a transmon capacitively coupled to a resonator and to a dedicated microwave drive line.
	The resonator possibly employs a Purcell filter which we do not include explicitly.
	In a frame rotating at the transmon-drive frequency~$\omega_d$ for both the resonator and the transmon, the Hamiltonian is time-independent and is given by
	\begin{align}
		H &= H_0 + H_c + H_d  \label{eq:H_full_main} \\
		H_0 &= \deltar \ar^\dagger \ar + \deltaq \bq^\dagger \bq + \frac{\anharm}{2}(\bq^\dagger)^2\bq^2 \label{eq:H0_main}\\
		H_c &= g(\ar\bq^\dagger + \ar^\dagger\bq) \label{eq:Hc_main}\\
		H_d &= \frac{\Omega}{2} (e^{i\phi}\bq+e^{-i\phi}\bq^\dagger) \label{eq:Hd_main}
	\end{align}
	where~$\ar$ and~$\bq$ are the creation operators for the resonator and the transmon, respectively; $\deltar=\omega_r-\omega_d$ and~$\deltaq=\omega_q-\omega_d$ with~$\omega_r$ and~$\omega_q$ the resonator and transmon frequencies, respectively; $\alpha<0$~is the transmon anharmonicity; $g$~corresponds to the capacitive coupling; $\Omega$~and~$\phi$ are the transmon-drive amplitude and phase, respectively.
	The phase is not relevant for the results in this work and we fix it to~$\phi=0$.
	
	We can qualitatively understand (see~\cref{fig1:readout_res_analytical}(a)) that~$H$ contains an effective coupling~$\tilde{g}$ between~$\ket{20}$ and~$\ket{01}$.
	If~$\omega_d$ matches the transition frequency between the \quotes{bare}~$\ket{20}$ and~$\ket{01}$, these two states are degenerate in the rotating frame and they are connected by two paths (at lowest order) via either~$\ket{11}$ or~$\ket{10}$.
	If~$\Delta\coloneqq\omega_q-\omega_r\gg g$ \addition{and}~$\deltaq\gg\Omega$, then~$\ket{11}$ and~$\ket{10}$ are occupied only \quotes{virtually} and one gets purely an effective $\ket{20}\leftrightarrow\ket{01}$ coupling.
	Modulo a constant term, in the 2D~subspace~$\drivesub=\linspan\{\ket{20},\ket{01}\}$ we can write~$H$ in~\cref{eq:H_full_main} as $H|_{\drivesub}\equiv-\eta(\omega_d)\ztype/2+\tilde{g}(\omega_d)\xtype$ for an appropriate function~$\eta$ (an approximation can be extracted from~\cref{eq:H0_ddressed}).
	As a function of~$\omega_d$ this Hamiltonian gives rise to a $\ket{20}\leftrightarrow\ket{01}$ avoided crossing centered at a frequency~$\omega_d^*$ (see~\cref{fig1:readout_res_analytical}(b)) where~$\eta(\omega_d^*)=0$.
	The energy separation at the center of the avoided crossing is then~$2\tilde{g}(\omega_d^*)$.
	
	In order to quantitatively study the action of~$H$, we unitarily transform it using a Schrieffer-Wolff transformation~$e^S$~\cite{Schrieffer66,Bravyi11,Magesan20,Boissonneault09}.
	Let~$\{\ket{ij}_\dressed\}$ be the basis of eigenvectors of~$H_0+H_c$ (the transmon-resonator \quotes{dressed} basis).
	In the dispersive regime ($g\ll\Delta$), with respect to a 1st-order Schrieffer-Wolff transformation~$S_1$ in the perturbation parameter~$g/\Delta$, such that $e^{-S_1}\ket{ml}\approx\ket{ml}_\dressed$, we get (see~\cref{sec:approx_hamil})
	\begin{align}
		H^{\dressed} \coloneqq e^S H e^{-S} &\approx e^{S_1} H e^{-S_1} \label{eq:Hdressed_main_firstline}\\
		&= H_0^{\dressed} + H_{d1}^{\dressed} + H_{d2}^{\dressed}
		\label{eq:Hdressed_main}
	\end{align}
	with
	\begin{align}
		H_0^{\dressed} &= \Bigl(\deltar  - \sum_{m=0}^\infty \frac{g^2\Delta_{-1}}{\Delta_m \Delta_{m-1}} \ket{m}\bra{m} \Bigr) \ar^\dagger\ar\nonumber\\
		&+ \sum_{m=1}^\infty \Bigl( m\deltaq +\frac{\anharm}{2}m(m-1) + \frac{g^2m}{\Delta_{m-1}} \Bigr) \ket{m}\bra{m}
		\label{eq:H0dressed_main}
	\end{align}
	\begin{align}
		H_{d1}^{\dressed} &= \frac{\Omega e^{i\phi}}{2} \bq + \text{h.c.} \\
		H_{d2}^{\dressed} &= \frac{\Omega e^{i\phi}}{2} \Biggl( 
		\ar \sum_{m=0}^\infty \frac{g\Delta_{-1}}{\Delta_m \Delta_{m-1}} \ket{m}\bra{m} \nonumber\\
		&\quad+ \ar^\dagger \sum_{m=0}^\infty \frac{g\anharm \sqrt{m+1}\sqrt{m+2}}{\Delta_m \Delta_{m+1}} \ket{m}\bra{m+2}
		\Biggr) + \text{h.c.},
		\label{eq;HDd2_main}
	\end{align}
	where~$\Delta_m\coloneqq \Delta +\anharm m$ and~$\{\ket{m}\}$ are transmon states.
	$H_0^{\dressed}$~is diagonal and contains the dispersive shifts, $H_{d1}^{\dressed}$ is the transmon drive now in the unitarily transformed frame, $H_{d2}^{\dressed}$~contains an indirect resonator drive and couplings of the kind~$\ar^\dagger \ket{m}\bra{m+2}$.
	In particular, for~$m=0$ in~\cref{eq;HDd2_main} we get a lowest order approximation of~$\tilde{g}$:
	\begin{align}
		\tilde{g} \approx \frac{\Omega g \anharm}{\sqrt{2}\Delta(\Delta+\anharm)}.
		\label{eq:tilde_g_lowest_main}
	\end{align}
	Notice that at this order there is no dependence on~$\omega_d$.
	Furthermore, $\tilde{g}$~would vanish for~$\anharm=0$, since the two paths in~\cref{fig1:readout_res_analytical}(a) fully destructively interfere in that case.
	Since~$\anharm$ is low for transmons, one can expect that~$\Omega$ needs to be relatively large for~$\tilde{g}$ to be substantial.
	
	For the drive to be most effective it is important that~$\omega_d$ matches~$\omega_d^*$.
	If~$g=0=\Omega$, there is no avoided crossing but~$\ket{20}$ and~$\ket{01}$ simply cross at $\omega_{d,0}^*\equiv 2\omega_q+\anharm -\omega_r$ as can be straightforwardly computed from~$H_0$ in~\cref{eq:H0_main}.
	This value is shifted due to the capacitive coupling (as can be seen from~\cref{eq:H0dressed_main}), as well as due to the possibly strong drive.
	For~$g\neq0$ and~$\Omega\neq0$ one can either compute~$\omega_d^*$ by full numerical diagonalization of~$H$ and find the avoided crossing as a function of~$\omega_d$, or one can find an (approximate) analytical expression.
	For the latter we use another Schrieffer-Wolff transformation (rather than the resolvent method in~\cite{Zeytinoglu15}, which does not give the full Hamiltonian) to account for the effect of the transmon drive~$H_{d1}^{\dressed}$ and to compute~$\omega_d^*$ up to order~$\Omega^4/(\deltaq)^3$, see~\cref{sec:approx_hamil}.
	We also use this transformation to compute~$\tilde{g}$ up to order~$\Omega^3/(\deltaq)^2$.
	Figures~\ref{fig1:readout_res_analytical}(c),(e) compare the analytical approach with the exact numerical results for~$\Delta\omega_d^*=\omega_d^*-\omega_{d,0}^*$ and~$\tilde{g}(\omega_d^*)$, respectively, given the parameters in~\cref{tab:target_parameters}.
	We consider 6~energy levels for the transmon and 3~for the resonator as we see that the exact curves converge for such choice.
	In~\cref{fig1:readout_res_analytical}(c)(d) we see that the two approximations are both pretty good, while in~\cref{fig1:readout_res_analytical}(e)(f) we see that~\cref{eq:tilde_g_lowest_main} \addition{deviates by up to~$1~\MHz$ from the exact value at high~$\Omega$ and that the absolute error with respect to the exact~$\tilde{g}(\omega_d^*)$ scales in a seemingly quadratic way.
	Instead, the higher order approximation stays closer to the exact curve and the error scales linearly.
	We expect that the remaining gap would be mostly filled by considering also higher orders in~$g/\Delta$ in the first Schrieffer-Wolff transformation, since increasing only the order of approximation in~$\Omega/\deltaq$ does not provide a significant improvement in~\cref{fig1:readout_res_analytical}(d).
	}
	
	\begin{table}
		\begin{tabular}{|c|c|c|}
			\hline
			\textbf{Parameter}                     &  Transmon     & Readout resonator                 \\
			\hline \hline
			Frequency {$\freq/2\pi$}      &     $6.7~\GHz$               &  $7.8~\GHz$ \\
			\hline
			Anharmonicity $\anharm/2\pi$    &     $-300~\MHz$                &  n.a. \\
			\hline
			Coupling $\barecoupling/2\pi$     &     \multicolumn{2}{c|}{$135~\MHz\quad\quad$} \\
			\hline
			Avg. photon number~$\avgphoton$     &   n.a.  &    0.005 \\
			\hline
			Relaxation time~$\Tone$    &   $30~\us$  &	$\addition{16~\ns}$ 	 	 \\
			&    &     ($\kappa/2\pi=10~\MHz$)   \\
			\hline
			Dephasing time~$\Ttwo$     &   $30~\us$  &    $\addition{32~\ns}$ \\
			&  (flux noise)  &    \\
			\hline
		\end{tabular}
		\caption{
			\label{tab:target_parameters}
			Parameters used both in the analysis and Lindblad simulations of the readout-resonator LRU, similar to the experimental ones in~\cite{Bultink19}.
			The transmon parameters correspond to the target parameters of a high-frequency data qubit in~\cref{sec:s17}.
		}
	\end{table}
	
	\subsection{Performance of the readout-resonator LRU}
	\label{sub:sims_readres}
	
	\begin{figure}
		\centering
		\includegraphics[width=\columnwidth]{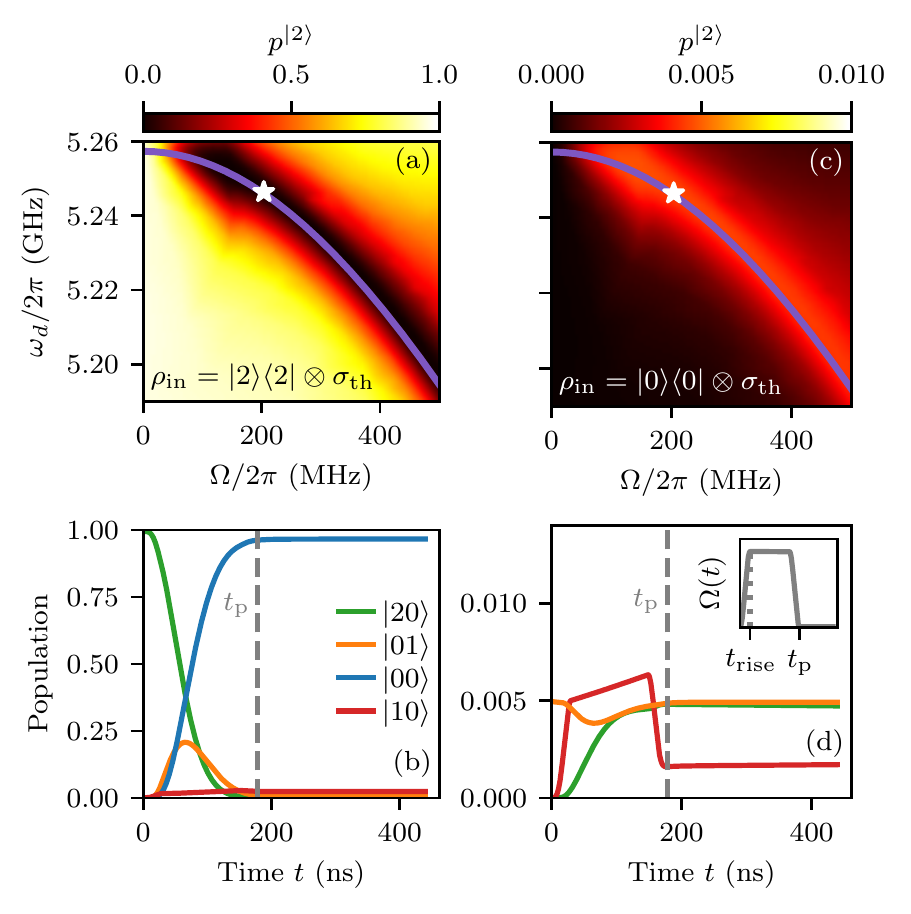}
		\caption{\label{fig2:readout_res_sims}
			Lindblad simulations of the transmon-resonator system for the readout-resonator LRU.
			In~(a),(b) the initial state is~$\ket{2}\bra{2}\otimes\thermal$, while in~(c),(d) it is~$\ket{0}\bra{0}\otimes\thermal$, where~$\thermal$ is the resonator thermal state.
			(a),(c)~Transmon leakage population~$\leakpop=\braket{2|\tr_r(\rho(\Tslot))|2}$ at the end of the time slot of~$\Tslot=440~\ns$.
			For each choice of~$(\Omega,\omega_d)$ we optimize the total pulse duration~$\tpulse\le\Tslot$ to minimize~$\leakpop$ given the initial state~$\ket{2}\bra{2}\otimes\thermal$, for fixed~$\trise=30~\ns$.
			The white star indicates the chosen operating point \addition{($\Omega/2\pi\approx204~\MHz$, $\omega_d/2\pi\approx5.2464~\GHz$, $\tpulse=178.6~\ns$) with~$\leakpop_\mathrm{op.}\approx0.5\%$ in~(a)}.
			The induced leakage in~(c) is~\addition{$\leakpop\approx0.48\%$} at the operating point.
			The purple line corresponds to the higher order estimate of the optimal drive frequency~$\omega_d^*$ as a function of~$\Omega$ (see~\cref{fig1:readout_res_analytical}(c)).
			The heatmaps are sampled using the \emph{adaptive} package~\cite{Nijholt19}.
			(b),(d)~Time evolution of the populations in a few selected states for the operating point.
			The vertical dashed line indicates the used~$\tpulse$.
			The inset in~(d) shows a schematic of the pulse~$\Omega(t)$.
		}
	\end{figure}
	
	Given the theoretical understanding of the transmon-resonator system, we devise a pulse to minimize the population in~$\ket{2}$ on a leaked transmon.
	We consider the pulse shape
	\begin{align}
		\pulse(t) = \begin{cases}
			\Omega \, \sin^2(\pi \frac{t}{2t_\mathrm{rise}}) \qquad&\mathrm{for}\quad 0\le t\le t_\mathrm{rise} \\
			\Omega \qquad&\mathrm{for}\quad t_\mathrm{rise} \le t\le \tpulse-t_\mathrm{rise} \\
			\Omega \, \sin^2(\pi \frac{\tpulse-t}{2t_\mathrm{rise}}) \qquad&\mathrm{for}\quad \tpulse-t_\mathrm{rise}\le t\le \tpulse
		\end{cases}
	\end{align}
	similarly to~\cite{Zeytinoglu15}, where~$\tpulse$ is the total pulse duration, at a fixed frequency~$\omega_d(t)=\omega_d$.
	Hence, there are four parameters to optimize over, i.e.~$\Omega,\omega_d,\tpulse$ and~$t_\mathrm{rise}$.
	We fix~$t_\mathrm{rise}=30~\ns$ since we observe that this strongly suppresses non-adiabatic transitions out of the manifold of interest: \addition{for example, $\ket{20}$ is coupled to~$\ket{10}$ by the drive but they are quite off-resonant, so only a fast pulse can cause \quotes{non-virtual} transitions between them.
	Indeed, for~$\trise\lesssim 10~\ns$ there appear ripples \addition{(for an example see~\cite{Zeytinoglu15})}} in~e.g.~the~$\ket{20}$ and~$\ket{10}$ populations when the drive is turned on and off, leading to a reduction in performance.
	We expect that an improved pulse shape can shorten~$t_\mathrm{rise}$.
	However, we do not explore this given the long maximum~$\tpulse$ allowed in our surface-code scheme~($\tpulse\le \Tslot=440~\ns$, see~\cref{sub:s17_layout}).
	
	We use Lindblad simulations of the transmon-resonator system to optimize over~$\Omega,\omega_d$ and~$\tpulse$.
	The Lindblad equation is given by
	\begin{align}
		\dot{\rho} = -i\bbrkt{H^{\dressed},\rho} + \sum_j \bigl( K_j\rho K_j^\dagger -\frac{1}{2} \{K_j^\dagger K_j, \rho\}\bigr)
		\label{eq:Lindblad}
	\end{align}
	with~$\{K_j\}$ the quantum jump operators.
	We express (and solve) this equation in the \addition{exact} unitarily transformed frame.
	\addition{That is, while in~\cref{sub:analytical_results_main} we have used a first-order Schrieffer-Wolff transformation~$e^{S_1}$ (see~\cref{eq:Hdressed_main_firstline}), in the numerics we compute the full transformation~$e^{S}$ (see also~\cref{eq:Hdressed_main_firstline}).
	In this way we find the basis that exactly diagonalizes~$H_0+H_c$ and express~$H_d$ in this basis as well, without any further Schrieffer-Wolff transformation like in~\cref{sub:analytical_results_main}.
	In other words, the simulations reproduce the dynamics under the Hamiltonian in~\cref{eq:H_full_main,eq:H0_main,eq:Hc_main,eq:Hd_main} without any approximation.}
	
	The Hamiltonian parameters are the same as in~\cref{sub:analytical_results_main} and are reported in~\cref{tab:target_parameters}, including the noise parameters.
	In particular, while we neglect the transmon thermal population, we include it for the resonator since it determines the leakage that the pulse induces when the transmon was not leaked, as we discuss below.
	The resonator thermal state is given by~\cite{BreuerPetruccione}
	\begin{align}
		\thermal \approx \Bigl( 1-\frac{\avgphoton}{1+2\avgphoton} \Bigr) \ket{0}\bra{0} + \frac{\avgphoton}{1+2\avgphoton} \ket{1}\bra{1}
	\end{align}
	for low average photon number~$\avgphoton$.
	We consider dressed relaxation and dephasing, as given below, assuming that this is a good model in the dispersive regime.
	In the unitarily rotated frame, the employed jump operators~$\{K_j\}$ are explicitly given by
	\begin{gather}
		\frac{1}{\sqrt{\Tone^r}}\ar=\addition{\sqrt{\kappa}}\ar,  \quad \sqrt{\frac{\avgphoton}{1+\avgphoton}}\addition{\sqrt{\kappa}} \ar^\dagger, \quad \sqrt{\frac{2}{\Tphi^r}}\ar^\dagger \ar,  \\
		\frac{1}{\sqrt{\Tone^q}}\bq, \quad \sqrt{\frac{2}{\Tphi^q}}\bq^\dagger \bq,
	\end{gather}
	where~$\Tphi=(1/\Ttwo-1/2\Tone)^{-1}$ \addition{and where we consider 6~energy levels for the transmon and 3~for the resonator}.
	Note that e.g.~for~$\ar$, going back to the original frame it holds that~$e^{-S}\ar e^{S}=  \sum_{l=0}^{1}\sqrt{l+1}\ket{l}_\dressed\bra{l+1}_\dressed=\ar_\dressed$ by definition of~$e^S$, corresponding indeed to relaxation in the dressed basis.
	By considering dressed relaxation and dephasing, the effective relaxation time~$\Tone^q$ of the transmon is not shortened by the fact that it is coupled to a lossy resonator (Purcell effect).
	\addition{We assume that this is a good approximation also during driving as the drive couples eigenstates which mostly have the same number of excitations in the resonator (except for~$\ket{20}$ and~$\ket{01}$ when the drive is near-resonant with this transition and causes a strong mixing of these states).}
	We thus mimic the use of a Purcell filter but without including it in the simulations since that would increase the Hilbert-space dimension in a computationally expensive way.
	
	For each choice of~$(\Omega,\omega_d)$ we optimize~$\tpulse$ such that, given the initial state~$\ket{2}\bra{2}\otimes \thermal$, the leakage population~$\leakpop=\braket{2|\tr_r(\rho(\Tslot))|2}$ at the end of the available time slot is minimized (see~\cref{fig2:readout_res_sims}(a)).
	\addition{The states~$\ket{20}$ and~$\ket{01}$ approximately form a two-level system with additional damping from~$\ket{01}$ to~$\ket{00}$, thus the drive effectively induces damped Rabi oscillations~\cite{Haroche06} between them.
	Oscillations occur only for~$\tilde{g}>\kappa/4$~\cite{Haroche06} (underdamped regime), while for~$\tilde{g}=\kappa/4$ (critical regime) or~$\tilde{g}<\kappa/4$ (overdamped regime) the populations in~$\ket{20}$ and~$\ket{01}$ simply decay in an exponential-like way without forming any minimum.
	For the parameters in~\cref{tab:target_parameters} the critical drive amplitude that gives~$\tilde{g}=\kappa/4$ is~$\Omega_\text{cr}/2\pi\approx143~\MHz$.
	Thus for~$\Omega\le\Omega_\text{cr}$ the best strategy is to drive until~$\leakpop$ reaches a (low) practically-stable value (which is in general not~0 when the full system is taken into account).
	Here with the given~$\kappa$ we find that this occurs in a time comparable to~$\Tslot$ only from about~$\Omega=\Omega_\text{cr}$, so for~$\Omega\le\Omega_\text{cr}$ we drive for the entire~$\Tslot$.
	For~$\Omega>\Omega_\text{cr}$ the optimization has many local minima as a function of~$\tpulse$, corresponding to the minima of the $\ket{20}\leftrightarrow\ket{01}$ oscillations induced by the drive. 
	Here we choose to target the first minimum as in~\cite{Zeytinoglu15,Egger18} since it is the fastest approach.
	For a sudden pulse this minimum would occur around~$\pi/2\tilde{g}$ for sufficiently small~$\kappa$, whereas we find heuristically that a good initial guess for the optimization is $\pi/2\tilde{g}_\text{damp}$ with~$\tilde{g}_\text{damp}\coloneqq\sqrt{\tilde{g}^2-(\kappa/4)^2}\,e^{-\kappa/7\tilde{g}}$ for larger~$\kappa$.
	Then for the optimization over~$\tpulse$ we use the bounds~$\tpulse-2\trise\in[0,1.1\times\pi/2\tilde{g}_\text{damp}]$ (using the bounded Brent method in~\emph{scipy}; we provide the code at~\url{https://doi.org/10.4121/14762052}).}
	While using a longer~$\tpulse$ in the underdamped regime (possibly even greater than the allotted~$\Tslot$) would eventually lead to an even lower leakage population~\cite{Magnard18}, it is not necessarily desirable as a longer~$\tpulse$ may mean that the disturbance to a non-leaked transmon is greater as well (see~\cref{sub:eth_approach}).
	
	\addition{While the procedure above optimizes~$\tpulse$ given a certain pair~$(\Omega,\omega_d)$, we use the package \emph{adaptive}~\cite{Nijholt19} to choose the next pair to sample and we iterate this process.
	This package searches a given parameter space (here~$\Omega/2\pi\in[0,500~\MHz]$, $\omega_d/2\pi\in[5.19,5.26~\GHz]$) in a finer way where the given cost function changes faster.
	Here we use~$(\log\leakpop)^2$ as the cost function since it changes faster where~$\leakpop$ is small, allowing us to get both a high-resolution heatmap (see~\cref{fig2:readout_res_sims}) and a good first estimation of the $\leakpop$~minima in a single run.
	Then we run a local optimization with tight bounds around some of these candidate points for fine tuning.}
	
	In~\cref{fig2:readout_res_sims}(a) one can observe a band with low~$\leakpop$ as desired.
	This band occurs at drive frequencies slightly \addition{above~$\omega_d^*(\Omega)$}, which one would expect to be optimal based on~\cref{sub:analytical_results_main}.
	We attribute this to the fact that a significant share of the time is taken by the rise and fall of the pulse, where~$\Omega(t)$ is smaller than the maximum.
	\addition{We find that one can choose a broad range of~$\Omega$s to achieve a~$\leakpop\gtrsim0.5\%$, from~$130~\MHz$ (slightly below the critical point) to deep in the underdamped regime.
	However, other considerations apply, namely, on the high end using a very high~$\Omega$ poses strong experimental requirements on the drive, while on the low end the pulse takes much longer and it is not a~priori given that driving at the critical point would be best.
	Actually, notice that driving at the critical point with good performance is possible only due to the relatively high~$\Tslot$ for the given~$\kappa$.}
	In the following we \addition{choose} the point marked by a star in~\cref{fig2:readout_res_sims} as the \addition{operating} point \addition{($\Omega/2\pi\approx204~\MHz$, $\omega_d/2\pi\approx5.2464~\GHz$, $\tpulse=178.6~\ns$)}.
	This point reaches~\addition{$\leakpop_\mathrm{op.}\approx0.5\%$} \addition{while affecting the least the coherence within the computational subspace (see~\cref{sub:more_heatmaps}).}
	We attribute the fact that this minimum does not reach~0 to \addition{re-heating from~$\ket{00}$ to~$\ket{01}$, as well as} transmon decoherence (resonator pure dephasing would contribute as well but here~$\Tphi^r=\infty$) \addition{and interactions with higher energy levels.}
	We note that in~\cref{fig2:readout_res_sims}(a) we find good~$\leakpop\lesssim5\%$ up to~$\Omega/2\pi\gtrsim\addition{100~\MHz}$, which could be used to further ease the requirements on the drive (see~\cref{sec:s17}).
	
	The time evolution for a few selected states is shown in~\cref{fig2:readout_res_sims}(b) for the operating point, given the initial state~$\ket{2}\bra{2}\otimes \thermal$.
	The first few~$\ns$ make~$\ket{20}$ rotate into~$\ket{01}$, while the latter decays relatively fast to~$\ket{00}$ due to the large relaxation rate~$\kappa$ of the readout resonator.
	\addition{Already after~$\approx220~\ns$ the remaining $\ket{01}$~population has practically returned to the thermal state.
	The repetition of the pulse, such as in the surface code (see~\cref{sec:s17}) at every QEC~cycle, thus does not lead to heating of the resonator with these system parameters (see~\cref{sec:discussion} for a discussion about other parameter regimes).}
	
	We now evaluate the effect of the pulse on a non-leaked transmon (see~\cref{fig2:readout_res_sims}(c),(d)).
	There should ideally be no effect, except for an acquired single-qubit phase which can easily be determined and corrected by either a real or virtual $Z$~rotation.
	First, if the transmon is in~$\ket{0}$ and there is some thermal population in the resonator, part of the state is supported on~$\ket{01}$, which rotates into~$\ket{20}$ in the same way as the opposite process by unitarity.
	\Cref{fig2:readout_res_sims}(c) shows that indeed the induced leakage is greater where~$\leakpop$ is lower in~\cref{fig2:readout_res_sims}(a).
	However, due to the low~$\avgphoton=0.005$, the induced leakage is also overall low~(\addition{$\leakpop\approx0.48\%$} in~\cref{fig2:readout_res_sims}(c) at the operating point, which is comparable to state-of-the-art $\CZ$~leakage rates, see~\cref{sub:lru_modeling_qsim}) and can be made even lower by engineering colder resonators.
	If the initial state is~$\ket{1}\bra{1}\otimes \thermal$ there is little induced leakage\addition{~($\leakpop\approx0.02\%$ at the operating point and $\leakpop\lesssim0.04\%$ across the whole landscape)} as the drive is off-resonant with transitions from this state.
	Second, the pulse might affect the coherence times of the transmon by driving transitions within or outside the computational subspace (and back), as the small but non-negligible transitory population in~$\ket{10}$ in~\cref{fig2:readout_res_sims}(b),(d) seems to suggest.
	However, we find that both the effective~$\Tone^q$ and~$\Ttwo^q$ are only marginally affected as a function of~$\Omega$ (see~\cref{sub:more_heatmaps}).
	This is because stronger pulses cause a somewhat stronger disturbance to the qubit, but they are shorter so that in total the effect is small.
	
	\section{Surface code with LRUs}
	\label{sec:s17}
	
	\subsection{Layout and operation scheduling}
	\label{sub:s17_layout}
	
	\begin{figure}
		\centering
		\includegraphics[width=\columnwidth]{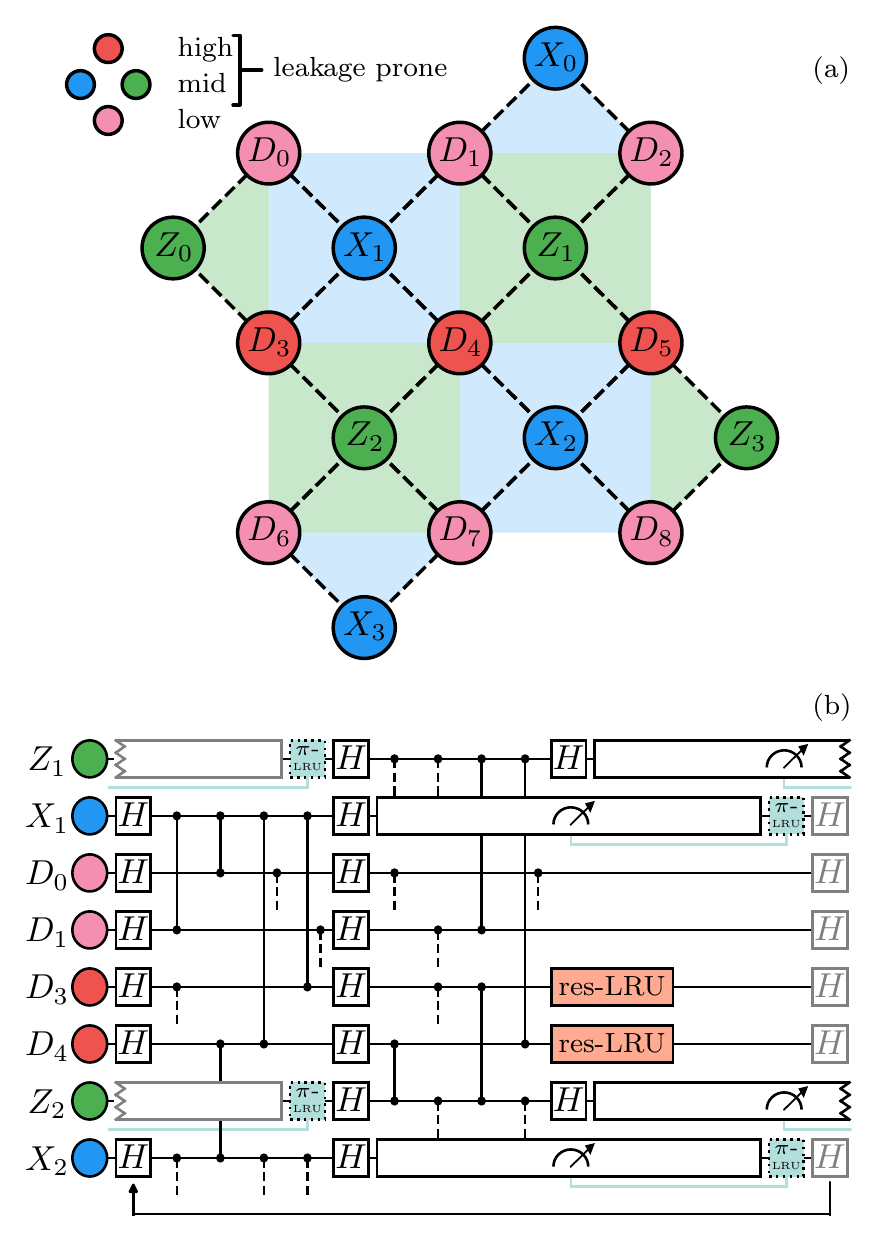}
		\caption{\label{fig3:s17}
			(a)~Schematic overview of the Surface-17 layout~\cite{Versluis17,Varbanov20}.
			Pink~(resp.~red) circles with \(\datatype\)~labels represent low- (high-) frequency data qubits, while blue (resp.~green) circles with \(\xtype\)~(\(\ztype\))~labels represent ancilla qubits, which have an intermediate frequency.
			Ancilla qubits and high-frequency data qubits are prone to leakage during the $\CZ$~gates.
			(b)~The quantum circuit for a single \QEC~cycle employed in simulation, for the unit-cell scheduling defined in~\cite{Versluis17}, in which we insert the LRUs.
			The \resLRU s~(orange) are applied unconditionally on the high-frequency data qubits after the~$\CZ$s, while the \piLRU s~(teal) are applied on the ancilla qubits depending on the measurement outcome.
			Gray elements correspond to operations belonging to the previous or the following \QEC~cycle.
			The duration of each operation is given in~\cref{sub:sim_params}.
			The arrow at the bottom indicates the repetition of \QEC~cycles.
		}
	\end{figure}
	
	We study the distance-3 rotated surface code (see~\cref{fig3:s17}(a)), nicknamed Surface-17, in the presence of leakage and LRUs.
	We follow the frequency and pipelined scheme in~\cite{Versluis17}, in which the 9~data qubits are subdivided into 3~high- and 6~low-frequency ones.
	The 4~$\xtype$ and the 4~$\ztype$ ancilla qubits have an intermediate frequency.
	We consider the flux-pulse implementation of the~$\CZ$s~\cite{Strauch03,DiCarlo09,Martinis14,Rol19a,Negirneac20} for tunable-frequency transmons, in which the transmon with the greater frequency is lowered towards the other one with a flux pulse.
	With this technique fluxed transmons are prone to leakage.
	This means that the high-frequency data qubits and all the ancilla qubits can leak.
	As shown in~\cite{Varbanov20}, leakage can last for many QEC~cycles and be quite detrimental to the logical performance of the code.
	Here we address these issues with the \resLRU~for high-frequency data qubits and with the \piLRU~for ancilla qubits, as described below.
	If due to a different implementation of the~$\CZ$s (or due to leakage mobility~\cite{Varbanov20,Mcewen21}) also the low-frequency data qubits can leak, one can apply the \resLRU~to them as well but we do not explore this here.
	
	The circuit executed for each QEC~cycle is shown in~\cref{fig3:s17}(b).
	The $\xtype$-type and $\ztype$-type parity-check units are implemented in an interleaved way, with the~$\CZ$s for one unit being applied while the other ancilla-qubit type is measured.
	The duration of each operation is summarized in~\cref{sub:sim_params}, with a total QEC-cycle duration of~$800~\ns$.
	The data qubits are idling for a considerable amount of time, namely~$\Tslot=440~\ns$, while the ancilla qubits are measured.
	We choose this time slot as the ideal place to apply the \resLRU s, introduced in~\cref{sec:readres}, to the high-frequency data qubits.
	Notice that the optimal pulse selected in~\cref{sub:sims_readres}, which was simulated for the target parameters of the high-frequency data qubits, takes about~$\tpulse=180~\ns$ \addition{and thus easily fits within this time slot (see~\cref{sec:discussion} for a discussion about other parameter regimes).}
	
	For the ancilla qubits there is no available time slot to apply the \resLRU.
	A possibility would be to make the QEC-cycle time longer by inserting these LRUs when the measurement is completed.
	However, this approach would lower the logical error rate of the code by a non-negligible amount.
	On the other hand, ancilla qubits are measured and the (analog) measurement outcome contains information about leakage~\cite{Varbanov20}.
	We choose to use a different type of LRU altogether which uses this information.
	Specifically, we consider a $\ket{1}\leftrightarrow\ket{2}$ $\pi$~pulse, conditioned on the measurement outcome reporting a~$\ket{2}$.
	Below we discuss further details of the implementation of this \piLRU.
	
	\subsection{Implementation of the LRUs in the density-matrix simulations}
	\label{sub:lru_modeling_qsim}
	
	We use density-matrix simulations~\cite{Obrien17} using the open-source package \emph{quantumsim}~\cite{quantumsim_website} to study Surface-17 with \resLRU s and~\piLRU s.
	We include relaxation and dephasing~($\Tone$ and~$\Ttwo$), as well as flux-dependent~$\Ttwo$ and leakage rate~$\leakrate$ during the~$\CZ$s, following the same error model as in~\cite{Varbanov20}.
	$\leakrate$~is defined as the average leakage from the computational to the leakage subspace~\cite{Wood18}.
	The state of the art is~$\leakrate\approx0.1\%$~\cite{Rol19a,Negirneac20}, although the actual~$\leakrate$ is expected to be higher when operating a multi-transmon processor~\cite{Krinner20,Marques21}, thus here we consider up to~$\leakrate=0.5\%$.
	We assume that single-qubit gates do not induce any leakage as their leakage rates are typically negligible compared to the~$\CZ$s~\cite{Chen16b,Babu20,Werninghaus20}.
	The noise parameters used are reported in~\cref{sub:sim_params}.
	Furthermore, during a~$\CZ$ with a leaked transmon, the non-leaked transmon acquires a phase called the leakage conditional phase~\cite{Varbanov20}.
	We select these phases \addition{uniformly} at random \addition{(see~\cref{sub:leakcondphases})} and, in contrast to~\cite{Varbanov20}, we then keep them fixed for every Surface-17 simulation in this work.
	This makes it easier to recognize trends as a function of the LRU~parameters.
	\addition{In~\cref{sub:leakcondphases} we discuss the variability of the logical error rate depending on the leakage conditional phases.}
	We do not consider further leakage from~$\ket{2}$ to~$\ket{3}$ in subsequent $\CZ$~gates~\cite{Varbanov20} as we expect it to be negligible when LRUs make~$\ket{2}$ short-lived.
	
	\subsubsection{\resLRU~for data qubits}
	\label{subsub:resLRU_model_qsim}
	
	In the simulations, leakage-prone transmons are modeled as 3-level systems and non-leakage-prone ones as 2-level systems, leading to an already computationally expensive size for the density matrix.
	As a consequence, we do not include the readout resonator explicitly in these simulations.
	The resonator is initially in the ground state and is returned to it at the end of the time slot, approximately.
	We can thus trace the resonator out and model the \resLRU~on the transmon qubit as an incoherent $\ket{2}\mapsto\ket{0}$~relaxation (\addition{see~\cref{subsub:resLRU_quantumsim}} for details).
	Furthermore, in~\cref{sub:sims_readres} we have observed that the \resLRU~can also cause a non-leaked transmon to partially leak, so we include that as an incoherent $\ket{0}\mapsto\ket{2}$~excitation.
	
	Calling~$p^{\ket{j}}_{i}, p^{\ket{j}}_{f}$ the populations before and after the \resLRU,  we define the leakage-reduction rate~$0\le\leakredrate\le 1$ as~$R = 1-\leakpop_{f}$ conditioned on an initially fully leaked transmon, i.e.~for~$\leakpop_{i}=1$.
	Furthermore, we define the average \resLRU~leakage rate~$\leakrateLRU$ as the average of the induced leakage starting from either~$\ket{0}$ or~$\ket{1}$ (consistently with the definition for~$\CZ$~\cite{Wood18}), with probability~$1/2$ each.
	Since almost all induced leakage comes from~$\ket{0}$ (see~\cref{sub:sims_readres}), this means that~$\leakpop_f\approx 0$ for~$\onepop_i=1$ and that~$\leakpop_f \approx 2\leakrateLRU$ for~$\zeropop_i=1$ (neglecting relaxation effects as the used~$\Tone=30~\us$ is relatively long).
	Combining these two definitions one gets the expression
	\begin{align}
		\leakpop_{f} \approx (1-R) \, \leakpop_{i} + 2\leakrateLRU \,\zeropop_{i}
	\end{align}
	for an arbitrary incoming state.
	Notice that, given these definitions, \cref{fig2:readout_res_sims}(a),(c) respectively show a heatmap of~$1-\leakredrate$ and~$2\leakrateLRU$ for the considered transmon-resonator parameters.
	In particular, the operating point achieves~\addition{$R\approx99.5\%$ and~$\leakrateLRU\approx0.25\%$}.
	The achieved leakage reduction can be compared with the one given purely by relaxation during~$\Tslot$, namely~$\leakredrate_{\Tone}=1-e^{-\Tslot/(\Tone/2)}=2.9\%$, which shows that the LRU provides a much stronger additional seepage channel.
	
	\subsubsection{\piLRU~for ancilla qubits}
	\label{subsub:piLRU_model_qsim}
	
	The dispersive readout of a transmon qubit is in general performed by sending a pulse to the readout resonator, integrating the reflected signal to obtain a point in the IQ~plane and depleting the photons in the resonator (either passively by relaxation or actively with another pulse)~\cite{Jeffrey14,Bultink16,Heinsoo18}.
	The measured point is compared to one or more thresholds to declare the measurement outcome.
	These thresholds are determined as to optimally separate the distributions for the different outcomes, which have a Gaussian(-like) form.
	Here we assume that the distribution for~$\ket{2}$ is sufficiently separated from~$\ket{0}$ and~$\ket{1}$~\cite{Jeffrey14}.
	This is generally expected to be possible thanks to the different dispersive shift.
	Then one uses three thresholds in the IQ~plane to distinguish between~$\ket{0}$, $\ket{1}$ and~$\ket{2}$ (or two if~$\ket{2}$ is well-separated from e.g.~$\ket{0}$).
	We also assume that an outcome can be declared during photon depletion, thus enabling real-time conditional feedback.
	This is challenging to perform in~200-300$~\ns$ in experiment due to the classical-postprocessing requirements, but it has been previously achieved~\cite{Riste12b,Andersen19}.
	We can then apply the \piLRU~right at the end of the depletion time.
	The $\ket{1}\leftrightarrow\ket{2}$ $\pi$~pulse is expected to be implementable as a simple pulse in the same way and time as single-qubit gates~($20~\ns$) and with comparable, coherence-limited fidelity.
	
	\addition{If conditional feedback is not possible in the allotted time, one can either increase the QEC-cycle duration (at the cost of extra decoherence for all qubits, scaling as~$1-e^{-t_\text{extra}/\Ttwo}$ per qubit per QEC~cycle) or postpone the conditional gate to the next QEC~cycle.
	In the latter case, one source of error corresponds to the ancilla qubit already seeping before the application of the~\piLRU, which then causes it to leak instead.
	The probability of this error is already low and is expected to become even lower with longer~$\Tone$s and lower-leakage~$\CZ$s.
	The other errors are the $\ztype$~rotations (depending on the leakage conditional phases) that the leaked ancilla qubit spreads for at least~1 extra QEC~cycle, as well as the fact that the parity-check stays disabled.
	We do not simulate these variants and we expect a relatively low logical-performance loss, corresponding to an average leakage lifetime of about 2~QEC~cycles (see~\cref{fig4:lifetime,fig:further_logerrrate}).}
	
	Readout-declaration errors are expected to affect the performance of the \piLRU.
	On one hand, an incorrect declaration of~$\ket{1}$ as a~$\ket{2}$ makes the $\pi$~pulse induce leakage.
	On the other hand, declaring a~$\ket{2}$ as a~$\ket{1}$ would lead to leakage not being corrected and lasting for at least one extra QEC~cycle.
	We define the readout matrix~$M$ with entries~$M_{ij}\eqqcolon \readoutp{i}{j}$ being the probability that the actual state~$\ket{j}$ resulting from the projective measurement is declared as an~$\ket{i}$. 
	In the simulations we use
	\begin{align}
		M = \begin{pmatrix}
			1  &  0  &   0 \\
			0  &  \readoutp{1}{1}  &   1- \readoutp{1}{1} \\
			0  &  1-\readoutp{2}{2}  &   \readoutp{2}{2} 
		\end{pmatrix}.
	\end{align}
	In particular, this means that we do not consider declaration errors within the computational subspace.
	While that would change the value of the logical error rate \addition{since the error syndrome gets corrupted}, it is not relevant for evaluating the performance of the \piLRU~\addition{since a~$\ket{0}$ mistaken for a~$\ket{1}$ or vice-versa does not trigger the \piLRU~anyway.}
	Furthermore, we assume that a~$\ket{0}$ cannot be mistaken as a~$\ket{2}$ since their readout signals are often much more separated than the signals of~$\ket{1}$ and~$\ket{2}$.
	\addition{Note that if a~$\ket{0}$ (rather than a~$\ket{1}$, as we assume in this work) could be mistakenly declared as a~$\ket{2}$, then a $\ket{1}\leftrightarrow\ket{2}$ $\pi$~pulse does not induce leakage, so here we consider the worst-case scenario for the \piLRU.}
	
	\subsection{Average leakage lifetime and steady state}
	\label{sub:lifetime}
	
	Once a qubit leaks, it tends to remain leaked for a significant amount of time, up to~10-15 QEC~cycles on average~\cite{Varbanov20}.
	Starting from an initial state with no leakage, the probability that a qubit is in the leaked state tends towards a steady state within a few QEC~cycles.
	It was shown in~\cite{Varbanov20} that this evolution is well captured by a classical Markov process with leakage (resp.~seepage) rate~$\GammaCL$~($\GammaLC$) per QEC~cycle, where~$\compsub$~(resp.~$\leaksub$) is the computational (leakage) subspace.
	Note that here~$\leaksub$ is 1-dimensional, corresponding to~$\ket{2}$.
	In our error model, without accounting for LRUs, these rates are approximately given by
	\begin{align}
		\GammaCL & \approx N_\flux\leakrate, \label{eq:gamma12_2state}                                    \\
		\GammaLC & \approx N_\flux\seeprate + (1-e^{-\frac{\tcycle}{\Tone/2}}), \label{eq:gamma21_2state}
	\end{align}
	where $N_\flux$~is in how many $\CZ$~gates the transmon is fluxed during a QEC~cycle, $\tcycle$~is the duration of a \QEC~cycle and~$\leakrate$~(resp.~$\seeprate$) is the average leakage (seepage) probability of a~$\CZ$~\cite{Wood18}.
	Thus the two native mechanisms that generate seepage are the~$\CZ$s themselves and relaxation.
	
	The major effect of a LRU is to effectively increase~$\GammaLC$ in~\cref{eq:gamma21_2state} by introducing an extra seepage mechanism.
	Hence we expect that~$\GammaLC^{\mathrm{LRU}}\sim \GammaLC + \leakredrate$ for data qubits and~$\GammaLC^{\mathrm{LRU}}\sim \GammaLC + \readoutp{2}{2}$ for ancilla qubits, preventing leakage from accumulating and lasting~long for large~$\leakredrate$ or~$\readoutp{2}{2}$.
	
	The average leakage lifetime~$\avglife$ is the average duration of leakage and for a Markov process it is calculated as
	\begin{align}
		\avglife &= \sum_{n=1}^{\infty} n \, \mathbb{P}(\mathrm{stay~in~} \mathcal{L} \mathrm{~for~} n \mathrm{~QEC~cycles}) \\
		&= \sum_{n=1}^{\infty} n (1-\GammaLC)^{n-1} \GammaLC = \frac{1}{\GammaLC},
	\end{align}
	thus assuming that the qubit starts in~$\mathcal{L}$.
	The evolution of the leakage probability~$\avgprob(\cycle)$, averaged over surface-code runs, as a function of the QEC-cycle number~$\cycle$ is well-approximated by~\cite{Varbanov20}
	\begin{equation}
		\avgprob(\cycle) = \frac{\Gamma_{\compsub\rightarrow \leaksub}}{\Gamma_{\compsub\rightarrow \leaksub}+\Gamma_{\leaksub\rightarrow \compsub}} (1-e^{-(\Gamma_{\compsub\rightarrow \leaksub}+\Gamma_{\leaksub\rightarrow \compsub})\cycle}).
		\label{eq:fit_pleak_evol}
	\end{equation}
	The steady state is the long-time limit and is given by
	\begin{align}
		\avgprob_\steady = \lim_{n\to\infty} \avgprob(n) = \frac{\Gamma_{\compsub\rightarrow \leaksub}}{\Gamma_{\compsub\rightarrow \leaksub}+\Gamma_{\leaksub\rightarrow \compsub}}.
	\end{align}
	For ancilla qubits $\avgprob(\cycle)$~can be computed directly from the \quotes{true} measurement outcomes (i.e.~without declaration errors on top).
	For data qubits it can be computed from the density matrix.
	Specifically, for data qubits we evaluate~$\avgprob(\cycle)$ right after the~$\CZ$s.
	
\begin{figure}
		\centering
		\includegraphics[width=\columnwidth]{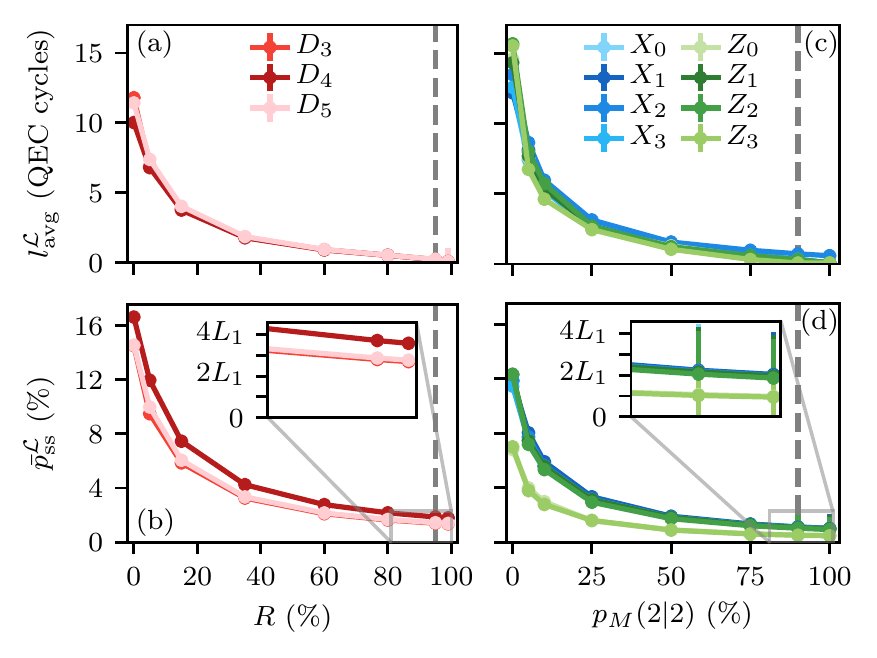}
		\caption{\label{fig4:lifetime}
			Average leakage lifetime~$\avglife$~[(a),(c)] and leakage steady state~$\avgprob_\steady$~[(b),(d)] as a function of the leakage-reduction rate~$\leakredrate$ for data qubits~[(a),(b)] and as a function of the readout probability~$\readoutp{2}{2}$ for ancilla qubits~[(c),(d)].
			Here we fix the $\CZ$~leakage rate to~$\leakrate=0.5\%$.
			The insets in~(b),(d) show that~$\avgprob_\steady$ tends to~$\approx N_\flux\leakrate$ ($N_\flux=4$ for~$\datatype_4$, 3~for~$\datatype_3,\datatype_5$, 1~for~$\ztype_0,\ztype_3$ and 2~for the remaining ancilla qubits).
			The vertical dashed lines correspond to the values used in~\cref{sub:logical_performance}.
			These results are extracted from $2\times10^4$~runs of 20~QEC~cycles each per choice of parameters.
			Error bars are estimated using bootstrapping and are mostly smaller than the symbol size.
		}
	\end{figure}

	\Cref{fig4:lifetime} shows~$\avglife$ and~$\avgprob_\steady$ extracted from the Surface-17 simulations by fitting~$\avgprob(\cycle)$ to~\cref{eq:fit_pleak_evol} for each qubit.
	We can indeed observe that these quantities drop substantially for both data and ancilla qubits.
	The decays follow an inverse proportionality as e.g.~for data qubits
	\begin{align}
		\avglife&=\frac{1}{\GammaLC^\mathrm{LRU}} \sim \frac{1}{\GammaLC+\leakredrate} \sim \frac{1}{\leakredrate} \\
		\avgprob_\steady&= \frac{\GammaCL^\mathrm{LRU}}{\GammaCL^\mathrm{LRU}+\GammaLC^\mathrm{LRU}} \sim \frac{\GammaCL^\mathrm{LRU}}{\GammaLC^\mathrm{LRU}}\sim\frac{\GammaCL^\mathrm{LRU}}{\leakredrate}
	\end{align}
	for sufficiently large~$\leakredrate$ and small~$\GammaCL^\mathrm{LRU}$.
	For ancilla qubits we expect, similarly, a~$1/\readoutp{2}{2}$ dependence.
	The lifetime drops from values~$\gtrsim 10$ to~$\approx1$, which is the minimum value it can achieve (some points drop below~1 within error bars as it is difficult for the fit to estimate such a short lifetime).
	As of course the LRUs do not prevent leakage from occurring during the~$\CZ$s in the first place, one cannot expect the steady state to reach~0 even for a perfect LRU~($\leakredrate=1$), but rather~$\avgprob_\steady\sim\GammaCL^\mathrm{LRU}\approx N_\flux\leakrate$ ($+\leakrateLRU$ if the LRU can mistakenly induce leakage).
	Figures~\ref{fig4:lifetime}(b),(d) show that this is indeed the case.
	
	\Cref{fig4:lifetime} also demonstrates that both~$\avglife$ and~$\avgprob_\steady$ get close to their minimum values already for~$\leakredrate,\readoutp{2}{2}\gtrsim 80\%$.
	This suggests that \resLRU~and \piLRU~may not necessarily need to be perfect to provide a good logical performance in Surface-17.
	This means that one could use e.g.~a weaker pulse to implement the \resLRU~or that the readout of~$\ket{2}$ may not need to be particularly optimized in practice.
	
	\subsection{Logical performance}
	\label{sub:logical_performance}
	
	In the simulations the logical qubit is initialized in~$\ket{0}_\mathrm{L}$ and the logical fidelity~$\logicalfid(\cycle)$ is computed at the end of each QEC~cycle as the probability that the decoder correctly determines \addition{whether a logical error has occurred or not.
	We do not perform a similar analysis with initial state~$\ket{+}_\mathrm{L}$ or other states as the density-matrix simulations are computationally expensive and we expect a similar performance.}
	The logical error rate~$\logerrrate$ per QEC~cycle can be extracted by fitting
	$\logicalfid(\cycle)=[1+\brkt{1-2\logerrrate}^{\cycle-\cycle_{0}}]/2$,
	where~\(\cycle_{0}\) is a fitting parameter (usually close to~0)~\cite{Obrien17}.
	We evaluate~$\logerrrate$ for the upper bound decoder~(\UB) which uses the complete density-matrix information to infer a logical error, and for the minimum-weight perfect-matching decoder~(\MWPM).
	\addition{Detailed information about these decoders can be found in~\cite{Obrien17,Obrien19} and an overview is given in~\cref{subsub:decoding}.}
	
	By mapping a leaked qubit back to the computational subspace, a LRU does not fully remove a leakage error but can at most convert it into a regular (Pauli) error.
	Hence, it is not to be expected that~$\logerrrate$ in the presence of leakage can be restored to the value at~$\leakrate=0$.
	We consider realistic parameters for the LRUs.
	Specifically, we use~$R=95\%$, $\leakrateLRU=0.25\%$, $\readoutp{2}{2}=90\%$ and~$\readoutp{1}{1}=99.5\%$.
	We have shown in~\cref{sub:sims_readres} that the first two parameters can be attained with realistic parameters for the transmon-readout system, while the last two are close to be  achievable in experiment~\cite{Magnard18,Sung20}.
	In particular, while the operating point has~\addition{$\leakredrate=99.5\%$}, we conservatively choose~$\leakredrate=95\%$ here.
	Notice that~$\readoutp{1}{1}=99.5\%$ is quite high.
	We argue that the state of the art can be squeezed as the threshold to distinguish between~$\ket{1}$ and~$\ket{2}$ in the IQ~plane could be moved towards~$\ket{2}$, rather than placing it in the middle as is common practice.
	In this way one would slightly reduce~$\readoutp{2}{2}$ in favor of~$\readoutp{1}{1}$ if~$\readoutp{1}{1}$ is not high enough.
	A broader study of the logical performance as a function of the LRU~parameters can be found in~\cref{sub:logerrrate_vs_LRUpars}.
	
\begin{figure}
		\centering
		\includegraphics[width=\columnwidth]{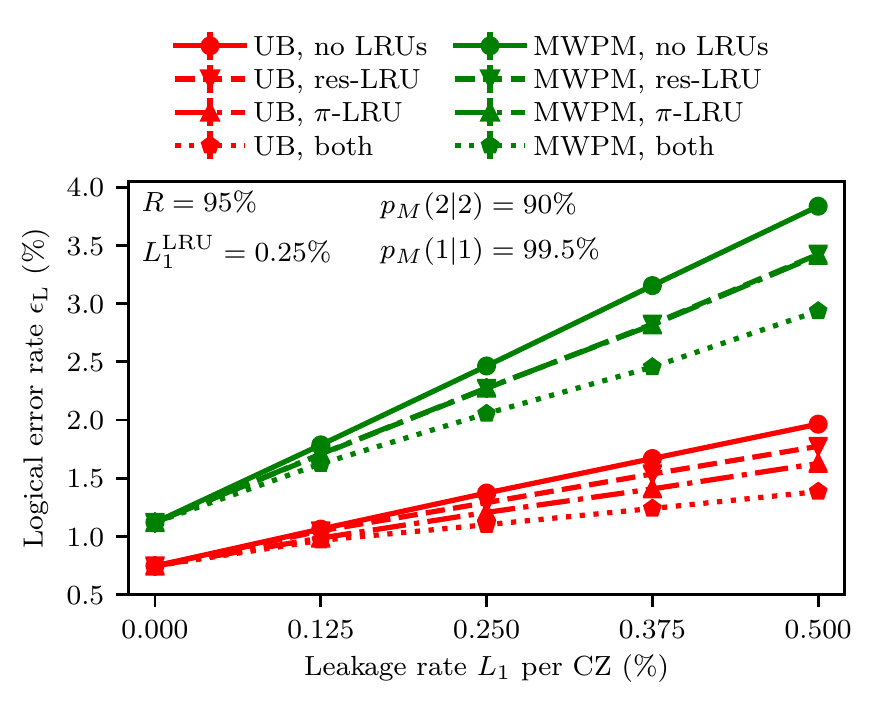}
		\caption{\label{fig5:logical_performance}
			Logical error rate~$\logerrrate$ per QEC~cycle for the upper bound~(\UB, red) and minimum-weight perfect-matching~(\MWPM, green) decoders versus the $\CZ$~leakage rate~$\leakrate$, in the cases with: no LRUs, only \resLRU, only \piLRU~and both LRUs (the point without leakage at~$\leakrate=0$ is always without LRUs as well).
			These results are extracted from $2\times10^4$~runs of 20~QEC~cycles each per choice of parameters.
			Error bars are estimated using bootstrapping and are smaller than the symbol size.
		}
	\end{figure}

	\Cref{fig5:logical_performance} shows the reduction in~$\logerrrate$ as a function of the $\CZ$~leakage rate~$\leakrate$ when LRUs with the given parameters are employed.
	Using only the \resLRU~or the \piLRU~lowers~$\logerrrate^\mathrm{\MWPM}$ by basically the same amount, while~$\logerrrate^\mathrm{\UB}$ is lower for the \piLRU~than for the \resLRU.
	We attribute this to the fact that \UB~directly uses the information in the density matrix, while \MWPM~relies on the measured syndrome, thus being more susceptible to ancilla-qubit leakage.
	When both LRUs are used, we see that~$\logerrrate$ is reduced by an amount which is close to the sum of the reductions when only one kind of LRU is used.
	As expected, $\logerrrate$~is not restored to the value at~$\leakrate=0$, but the reduction is overall significant and can reach up to~$30\%$ for both \MWPM~and \UB~compared to the case without LRUs.
	
	\section{Discussion}
	\label{sec:discussion}
	
	In this work we have introduced a leakage-reduction scheme using \resLRU s and \piLRU s which does not require any additional hardware or a longer QEC~cycle.
	Furthermore, while the scheme in~\cite{Mcewen21} is applicable only to ancilla qubits, our combination of \resLRU~for data qubits and \piLRU~for ancilla qubits enables to significantly reduce leakage in the whole transmon processor.
	We have shown with detailed simulations using realistic parameters that the reset scheme in~\cite{Zeytinoglu15,Egger18,Magnard18} can be adapted to be a LRU without significantly affecting the states in the computational subspace, allowing to unconditionally apply the \resLRU~in the surface code.
	The use of the \resLRU~for data qubits, as well as the use of the \piLRU~for ancilla qubits, leads to a substantial reduction of the average leakage lifetime and leakage steady state, preventing leakage from lasting more than~$\approx 1$~QEC~cycles on average, even when the LRUs are imperfect and can introduce leakage themselves.
	Using full density-matrix simulations of Surface-17 we have demonstrated that this leads to a significant reduction of the logical error rate for both the \UB~and \MWPM~decoders.
	
	Regarding the practical implementation of the \resLRU, the required drive amplitude is relatively strong, similarly to the one used in the experiments in~\cite{Zeytinoglu15,Egger18,Magnard18}.
	It is thus important that the microwave crosstalk is minimized by careful engineering of the drive lines.
	Furthermore, in a multi-transmon processor it is relevant that the drive frequency does not accidentally match any two-qubit or neighboring single-qubit transitions. 
	E.g.,~in the original scheme in~\cite{Versluis17} that we followed, the target frequencies are $6.7$,~$6.0$~and~$4.9~\GHz$ for high-, mid- and low-frequency qubits, respectively, and~$7.8~\GHz$ for the readout resonator~\cite{Bultink19}.
	In particular, the mid-frequency qubits (the ancilla qubits) are parked around~5.4-$5.5~\GHz$ during measurement, with their $\ket{1}\leftrightarrow\ket{2}$~transition around~5.1-$5.2~\GHz$.
	This is close to the optimal drive frequency found in~\cref{sub:sims_readres}~($\approx5.25~\GHz$), which can lead to an indirect ancilla-qubit drive mediated by the bus resonator, albeit weaker.
	The difficulty of precise frequency targeting in fabrication can further lead to undesired frequency collisions.
	These issues can be alleviated by choosing slightly different transmon/resonator frequencies and anharmonicities to make the drive more off-resonant with that transition (combined with better frequency targeting~\cite{Hertzberg20}), or they can be mitigated altogether by using tunable couplers~\cite{Yan18,Arute19,Sung20}.
	The \resLRU~is compatible with tunable-coupler schemes and their possibly different operation scheduling than in~\cite{Versluis17}, as well as potentially applicable to superconducting qubits which use a resonator for dispersive readout other than the transmon.
	\addition{Tunable couplers would also be advantageous to fully protect the \resLRU~performance from residual $ZZ$~crosstalk, even though we find that a cumulative $ZZ$~interaction up to~$\sim2~\MHz$ can be tolerated with fixed couplers (see~\cref{sub:residualZZ}).}
	Beside this, if the low-frequency data qubits can leak depending on the implementation of the~$\CZ$, the \resLRU~can be applied to them in the same time slot as the high-frequency ones.
	If the thermal population in the readout resonator is relatively high in a given experiment, the effect of a correspondingly high~$\leakrateLRU$ can potentially be mitigated by applying \resLRU~conditionally on the detection of leakage by a set of hidden Markov models~\cite{Varbanov20}.
	
	\addition{Regarding the viability of inserting the \resLRU~in the surface-code time scheduling, the necessary condition is that $\tpulse\le\Tslot$.
	We can express~$\Tslot$ as $\Tslot=\tmeas-4\tcz$, where~$\tmeas$ is the measurement time for the ancilla qubits.
	Slower~$\CZ$s might make~$\Tslot$ too short, although~$\CZ$s even faster than~$40~\ns$ (as assumed here) have been realized in~$15~\ns$~\cite{Foxen20}.
	The measurement time can be further broken down into readout-pulse time and photon-depletion time, $\tmeas=t_\mathrm{read}+t_\mathrm{depl}$.
	Both of these would be reduced by a larger~$\kappa$, however, assuming that the~$\kappa$'s of ancilla- and data-qubit resonators are comparable, $\tpulse$~would be reduced as well.
	Even if we keep~$\tpulse$ and~$\tcz$ fixed to the values in this work, we get~$\tmeas\ge340~\ns$, which is significantly lower than~$\tmeas=580~\ns$ as considered here.
	A desirable, additional condition to the necessary one is that $\Tslot-\tpulse\ge4/\kappa$, i.e.~that there is enough leftover time in~$\Tslot$ to allow for the data-qubit resonator to return the thermal state, where we estimate that 4~decay constants would suffice (together with the fact that the resonator was already relaxing during~$\tpulse$).
	Assuming similar depletion time for data- and ancilla-qubit resonators, this roughly means that the \resLRU~is easily applicable if~$\tpulse$ is smaller or similar to~$t_\mathrm{read}$.
	Note that in this work we have $\Tslot-\tpulse\sim16/\kappa$ and $\tpulse<t_\mathrm{read}$.
	If the additional condition above is not satisfied, one could demand that at least the resonator has returned to the thermal state before the \resLRU~in the following QEC~cycle, i.e.~$\Tslot-\tpulse+8\tcz+2t_\mathrm{H}\ge4/\kappa$.
	In this case the disadvantage would be that the presence of a fraction of a photon in the resonator would cause additional data-qubit dephasing especially during the first few~$\CZ$s.
	As the extra photon is present only when the qubit was previously leaked, we expect this disadvantage to be small as long as the overall leakage rate is small.
	If even the relaxed additional condition is violated, on top of the additional dephasing the resonator would also heat up, effectively leading to a higher~$\leakrateLRU$ in the QEC~cycle(s) following the one in which the qubit leaked.
	As also this effect scales with~$\leakrate$, we expect that it would not be an issue as long as~$\kappa$ is not very low (allowing for at most 1~extra QEC~cycle to thermalize we get~$\kappa/2\pi\ge1~\MHz$).
	Otherwise, leakage would not really be removed from the system but would be largely moved back and forth from the transmon to the resonator.}
	
	The demonstrated reduction in the average leakage lifetime and in the logical error rate is expected to lead to a higher noise threshold for the surface code in the presence of leakage, compared to the case without LRUs.
	\addition{Furthermore, for error rates below threshold (both regular and leakage) we believe that the logical error rate would be exponentially suppressed with increasing code distance when employing LRUs.
	Without LRUs this might hold only when the code distance is sufficiently larger than the average leakage lifetime~($\distance\gg\avglife$).
	For smaller distances the relatively long correlated error chains induced by leakage might lead to a sub-exponential scaling.}
	To study the noise threshold \addition{and sub-threshold behavior} it is necessary to implement simulations of large code sizes which use a simplified error model, such as a stochastic error model for leakage and Pauli errors~\cite{Fowler13,Kelly15,Suchara15}.
	We expect that the demonstrated \MWPM~logical error rate can be further lowered by the use of decoders~\cite{Fowler13,Kelly15,Suchara15,Stace10,Nagayama17,Auger17} that use information about leakage extracted directly or indirectly (e.g.~with hidden Markov models~\cite{Varbanov20}) from the measurement outcomes.
	
	\bigskip
	
	The data underlying this work, as well as the code to analyze it, are available at \url{https://doi.org/10.4121/c.5320331}.
	The code used to generate the data is available upon request to the corresponding author.
	
	\begin{acknowledgments}
		We thank L.~DiCarlo for his comments on the manuscript.
		F.B., B.M.V.~and B.M.T.~are supported by ERC grant EQEC No.~682726.
		Most simulations were performed with computing resources granted by RWTH Aachen University under projects rwth0566 and rwth0669.
		The data and code are available upon request from the corresponding author (f.battistel@tudelft.nl).
	\end{acknowledgments}

	\appendix
	
	\section{Approximate transmon-resonator Hamiltonian}
	\label{sec:approx_hamil}
	
	\subsection{Schrieffer-Wolff Transformation}
	\label{sub:SW}
	
	In this section we explain the concept of the Schrieffer-Wolff transformation~(\SW)~\cite{Schrieffer66,Bravyi11,Magesan20} and derive the equations that we use in the following sections.
	
	Consider a Hamiltonian
	\begin{equation}
		H=H_0+\epsilon V
	\end{equation}
	expressed in a certain basis~$\{\ket{\psi_n}\}$, where~$H_0$ is block diagonal with respect to this basis and the perturbation~$V$ can be taken as block off-diagonal without loss of generality (block-diagonal terms can be included in the definition of~$H_0$).
	Furthermore, we assume~$\norm{V}=\mathcal{O}\brkt{1}$ and~$\epsilon\ll\Delta_{ij}$, where we set~$\Delta_{ij}$ as the minimum energy separation between blocks~$i$ and~$j$.
	
	The \SW~corresponds to finding an anti-hermitian matrix~$S$ such that
	\begin{equation}
		H'\coloneqq e^S H e^{-S}
	\end{equation}
	is block diagonal.
	In other words, calling~$\{\ket{\bar{\psi}_n}\}$ the basis of eigenstates of~$H$, $e^S=\sum_n \ket{\psi_n} \bra{\bar{\psi}_n}$.
	The matrix~$S$ can be expanded in a series
	\begin{equation}
		S=\sum_{k=1}^{\infty} \epsilon^k S_k
	\end{equation}
	where each~$S_k$ is block off-diagonal.
	If~$\epsilon\ll \Delta_{ij}$ one can expect the first order~($S_1$) to provide a good approximation, otherwise one needs to consider higher orders depending on~$\epsilon$ (although the series does not always converge for extensive systems~\cite{Bravyi11}).
	Using the Baker-Campbell-Hausdorff formula one gets
	\begin{align}
		H' = e^S H &e^{-S} = \sum_{k=0}^{\infty} \frac{1}{k!} \underbrace{[S,[S,\dots[S,}_{k~\text{times}} H]\dots]].
		\label{eq:BCH}
	\end{align}
	The procedure for the \SW~is to group terms of the same order in~$\epsilon$ in this formula and set the block off-diagonal part of~$H'$ to~0, thus getting equations for~$\{S_k\}$, in the usual case with \emph{two} blocks~\cite{Bravyi11}.
	One uses the relationships
	\begin{align}
		&\bbrkt{\text{diagonal}, \text{diagonal}} = \text{diagonal},\\
		&\bbrkt{\text{diagonal}, \text{off-diagonal}} = \text{off-diagonal}, \\
		&\bbrkt{\text{off-diagonal}, \text{off-diagonal}} = \text{diagonal}.
	\end{align}
	However, the last line only holds for the case with two blocks.
	In the following we consider the generalization of the~\SW~to the case with an arbitrary number of blocks~\cite{Magesan20}.
	We use the notation~$O_{\diag}$ and~$O_{\offdiag}$ for the block diagonal and off-diagonal parts of an operator~$O=O_{\diag}+O_{\offdiag}$, respectively.
	
	Here we expand~$H$ and~$S$ up to~$k=3$ in~\cref{eq:BCH}, assuming that the 4th-order block off-diagonal term is negligible.
	We get the following pieces:
	\begin{widetext}
		\begin{align}
			\text{0th order}: \qquad &H_0 \\
			\text{1st order}: \qquad &V + \bbrkt{S_1,H_0} \\
			\text{2nd order}: \qquad & \bbrkt{S_1,V} + \frac{1}{2} \bbrkt{S_1, \bbrkt{S_1,H_0}} + \bbrkt{S_2,H_0}\\
			\text{3rd order}: \qquad & \bbrkt{S_2,V} + \frac{1}{2} \Bigl(\bbrkt{S_2, \bbrkt{S_1,H_0}} +  \bbrkt{S_1, \bbrkt{S_1,V}}
			+ \bbrkt{S_1, \bbrkt{S_2,H_0}}\Bigr) \nonumber\\ 
			&+ \frac{1}{6} \bbrkt{S_1, \bbrkt{S_1, \bbrkt{S_1,H_0}}} + \bbrkt{S_3,H_0}\\
			\text{4th order}: \qquad & \bbrkt{S_3,V} + \frac{1}{2} \Bigl(
			\bbrkt{S_1, \bbrkt{S_3,H_0}} + \bbrkt{S_2, \bbrkt{S_2,H_0}}
			+ \bbrkt{S_3, \bbrkt{S_1,H_0}} 
			+ \bbrkt{S_1, \bbrkt{S_2,V}} + \bbrkt{S_2, \bbrkt{S_1,V}}
			\Bigr) \nonumber\\
			&+ \frac{1}{6} \Bigl(
			\bbrkt{S_1, \bbrkt{S_1, \bbrkt{S_1,V}}} 
			+ \bbrkt{S_2, \bbrkt{S_1, \bbrkt{S_1,H_0}}} + \bbrkt{S_1, \bbrkt{S_2, \bbrkt{S_1,H_0}}} 
			+ \bbrkt{S_1, \bbrkt{S_1, \bbrkt{S_2,H_0}}}
			\Bigr) \nonumber\\
			&+ \frac{1}{24} \bbrkt{S_1, \bbrkt{S_1, \bbrkt{S_1, \bbrkt{S_1,H_0}}}}.
		\end{align}
	\end{widetext}
	Setting the block off-diagonal parts at 1st, 2nd and 3rd order to~0 we get
	\begin{align}
		\bbrkt{H_0,S_1} &= V \label{eq:SW_S1_defeq}\\
		\bbrkt{H_0,S_2} &= \frac{1}{2}\bbrkt{S_1,V}_\offdiag \label{eq:SW_S2_defeq}\\
		\bbrkt{H_0,S_3} &= \frac{1}{2}\bbrkt{S_2,V}_\offdiag + \frac{1}{3} \bbrkt{S_1, \bbrkt{S_1,V}_\diag}_\offdiag \nonumber\\
		&\quad + \frac{1}{12} \bbrkt{S_1, \bbrkt{S_1,V}_\offdiag}_\offdiag,
		\label{eq:SW_S3_defeq}
	\end{align}
	where we have used the first equation to simplify the following ones.
	These equations can be solved iteratively for~$S_k$ (given knowledge of the eigenstates of~$H_0$).
	The Hamiltonian~$H'$ is then block diagonal up to 4th~order and is explicitly given by
	\begin{align}
		H' = &H_0 + \frac{\epsilon^2}{2}\bbrkt{S_1,V}_\diag \nonumber\\
		&+ \epsilon^3 \Bigl(\frac{1}{2}\bbrkt{S_2,V}_\diag + \frac{1}{12} \bbrkt{S_1, \bbrkt{S_1,V}_\offdiag}_\diag\Bigr) \nonumber\\
		&+\epsilon^4 \Bigl( \frac{1}{2} \bbrkt{S_3,V}_\diag -\frac{1}{24} \bbrkt{S_1,\bbrkt{S_1,\bbrkt{S_1,V}_\diag}_\offdiag}_\diag
		\nonumber\\
		&\qquad-\frac{1}{6} \bbrkt{S_2,\bbrkt{S_1,V}_\offdiag}_\diag  +\frac{1}{12} \bbrkt{S_1,\bbrkt{S_2,V}_\offdiag}_\diag
		\Bigr).
		\label{eq:SWT_hamil_diag}
	\end{align}
	This expression has been simplified using~\cref{eq:SW_S1_defeq,eq:SW_S2_defeq,eq:SW_S3_defeq}, together with the fact that e.g.~$\bbrkt{S_k,\bbrkt{\ldots,\ldots}_\diag}_\diag=0$ since~$S_k$ is block off-diagonal.
	
	\subsection{\SW~of the capacitive coupling}
	\label{sub:SW_JC}
	
	We consider the Hamiltonian~$H = H_0 + H_c + H_d$ of a driven transmon capacitively coupled to a resonator, as given in~\cref{eq:H_full_main,eq:H0_main,eq:Hc_main,eq:Hd_main}.
	
	The \SW~of~$H_c$ up to 1st~order in the perturbation parameter~$\epsilon=g/\Delta$, where~$\Delta=\omega_q-\omega_r$, is implemented using the matrix~\cite{Boissonneault09}
	\begin{align}
		S_1 = g\sum_{m=1}^\infty \frac{\sqrt{m}}{\Delta+\anharm(m-1)} \Bigl(
		\ar\ket{m}\bra{m-1}  - \text{h.c.}
		\Bigr),
	\end{align}
	where~$\{\ket{m}\}$ are transmon states and where we have absorbed~$\epsilon$ in the definition of~$S_1$.
	The Hamiltonian in the unitarily transformed frame as defined in~\cref{sub:analytical_results_main} is then given by
	\begin{align}
		H^\dressed \approx e^{S_1} H e^{-S_1} =  e^{S_1} (H_0 + H_c) e^{-S_1} + e^{S_1} H_d \, e^{-S_1}
	\end{align}
	with
	\begin{widetext}
		\begin{align}
			e^{S_1} (H_0 + H_c) e^{-S_1} &= H_0 + \frac{1}{2} \bbrkt{S_1,H_c} \\
			&\approx \deltar \ar^\dagger\ar + \sum_{m=1}^\infty \Bigl( m\deltaq +\frac{\anharm}{2}m(m-1) + \frac{g^2m}{\Delta_{m-1}} \Bigr) \ket{m}\bra{m}
			-\ar^\dagger\ar \sum_{m=0}^\infty \frac{g^2\Delta_{-1}}{\Delta_m \Delta_{m-1}} \ket{m}\bra{m}
			\label{eq:dressed_hamil_non_driven} \\
			&\coloneqq H_{0}^{\dressed}
		\end{align}
	\end{widetext}
	where we define $\Delta_m = \Delta+\anharm m = \Delta-\abs{\anharm} m$ as~$\anharm<0$ for transmons.
	The second term above contains a Stark shift of the transmon frequency and the last term is the state-dependent dispersive shift.
	The approximation in~\cref{eq:dressed_hamil_non_driven} is due to the fact that we have ignored a double-excitation exchange term coming from~$\bbrkt{S_1,H_c}$, since it is proportional to~$g\alpha/(\Delta_m\Delta_{m-1})$.
	This is negligible for low anharmonicity and, secondly, for~$\omega_r>\omega_q$ as then~$\Delta<0$ and~$\abs{\Delta_m}$ increases with~$m$.
	If instead~$\omega_r<\omega_q$, $\Delta>0$ and~$\abs{\Delta_m}$~decreases with~$m$, so even if the approximation is good for the two lowest levels, there can be some higher level which does not sit well within the dispersive regime.
	However, in this work we consider a system with~$\omega_r>\omega_q$, hence we do not need to take this into account.
	
	The drive Hamiltonian in the unitarily transformed frame takes the form
	\begin{widetext}
		\begin{align}
			e^{S_1} H_d \, e^{-S_1} = \underbrace{\frac{\Omega e^{i\phi}}{2} \bq + \text{h.c.}}_{\coloneqq H_{d1}^{\dressed}} +  \underbrace{\frac{\Omega e^{i\phi}}{2} \Biggl( 
				\ar \sum_{m=0}^\infty \frac{g\Delta_{-1}}{\Delta_m \Delta_{m-1}} \ket{m}\bra{m}
				+ \ar^\dagger \sum_{m=0}^\infty \frac{g\anharm \sqrt{m+1}\sqrt{m+2}}{\Delta_m \Delta_{m+1}} \ket{m}\bra{m+2}
				\Biggr) + \text{h.c.}}_{\coloneqq H_{d2}^{\dressed}}
			\label{eq:dressed_drive_hamil}
		\end{align}
	\end{widetext}
	The last term contains a 1st-order approximation in~$g/\Delta$ of the $\ket{20}\leftrightarrow\ket{01}$ effective coupling~$\tilde{g}$, which is linear in~$\Omega$.
	However, the \quotes{pure} drive term~$H_{d1}^{\dressed}$ can be quite strong, so we need to evaluate how it affects~$\tilde{g}$ and the rest of the Hamiltonian.
	
	\subsection{\SW~of the pure drive Hamiltonian}
	\label{sub:SW_drive}
	
	Summarizing, in the unitarily transformed frame the original Hamiltonian~$H$ takes (approximately) the form
	\begin{align}
		H^D \approx H_0^D + H_{d1}^{\dressed} +  H_{d2}^{\dressed},
	\end{align}
	where~$H_0^D$ is given in~\cref{eq:dressed_hamil_non_driven} and~$H_{d1}^{\dressed}, H_{d2}^{\dressed}$ are given in~\cref{eq:dressed_drive_hamil}.
	
	We now want to find an additional \SW~transformation~$S'=S_1'+S_2'+S_3'$, with~$H_{d1}^{\dressed}$ taking the role of~$V$ in~\cref{sub:SW}, defining a \quotes{double-dressed} Hamiltonian
	\begin{align}
		H^{\ddressed} &\coloneqq e^{S'} H^\dressed e^{-S'} \\
		&= \underbrace{e^{S'} (H_0^\dressed + H_{d1}^{\dressed}) e^{-S'}}_{\eqqcolon H_0^{\ddressed}} + \underbrace{e^{S'} H_{d2}^{\dressed} \, e^{-S'}}_{\eqqcolon H_d^{\ddressed}}
		\label{eq:H_ddressed_full}
	\end{align}
	such that~$H_0^{\ddressed}$ is fully diagonal up to 3rd~order in the perturbation parameter~$\epsilon=\Omega/\deltaq$.
	Then~$H_d^{\ddressed}$ gives the couplings within the manifold of interest ($\ket{20},\ket{01}$) and outside of it.
	We absorb~$\epsilon^k$ in the definition of~$S_k'$ so it does not explicitly appear below.
	
	Following~\cref{sub:SW}, to find~$S_1'$ we need to solve~\cref{eq:SW_S1_defeq}, i.e.
	\begin{align}
		\bbrkt{H_0^\dressed, S_1'} = H_{d1}^{\dressed}
	\end{align}
	in this specific case.
	Bracketing it with the eigenstates~$\{\ket{ml}\}$ of~$H_0^\dressed$, with the notation $\ket{\mathrm{transmon,resonator}}$, we get the matrix elements of~$S_1'$ as
	\begin{align}
		\braket{ml|S_1'|nk} = \frac{\braket{ml|H_{d1}^{\dressed}|nk}}{E_{ml}^{\dressed}-E_{nk}^{\dressed}},
	\end{align}
	where~$\{E_{ml}^{\dressed}\}$ are the eigenenergies of~$H_0^\dressed$, which can be easily inferred from~\cref{eq:dressed_hamil_non_driven}.
	We neglect the dispersive shift since it is proportional to~$\anharm/\Delta$.
	Then
	\begin{align}
		\braket{ml|S_1'|nk} = &\frac{\Omega}{2} \Biggl(
		- \frac{\sqrt{m+1}\delta_{m,n-1}\delta_{l,k}}{\deltaq+\anharm m+\frac{g^2\Delta_{-1}}{\Delta_{m-1}\Delta_m}} e^{i\phi} \nonumber\\
		&\qquad + \frac{\sqrt{m}\delta_{m,n+1}\delta_{l,k}}{\deltaq+\anharm (m-1)+\frac{g^2\Delta_{-1}}{\Delta_{m-2}\Delta_{m-1}}} e^{-i\phi}
		\Biggr),
	\end{align}
	where~$\delta_{i,j}$ is the Kronecker delta.
	From this equation one can infer that
	\begin{align}
		S_1' = -\frac{\Omega}{2}e^{i\phi} \sum_{m=0}^\infty \frac{\sqrt{m+1}}{\deltaq_m} \ket{m}\bra{m+1} - \text{h.c.},
		\label{eq:S1prime}
	\end{align}
	where we have defined~$	\deltaq_m = \deltaq+\anharm m+\frac{g^2\Delta_{-1}}{\Delta_{m-1}\Delta_m}$.
	
	Having derived~$S_1'$, we can compute~$S_2'$ from~\cref{eq:SW_S2_defeq}, i.e.
	\begin{align}
		\bbrkt{H_0^{\dressed},S_2'} = \frac{1}{2}\bbrkt{S_1',H_{d1}^{\dressed}}_\offdiag
	\end{align}
	with
	\begin{align}
		\bbrkt{S_1',H_{d1}^{\dressed}} = &- \frac{\Omega^2}{2} \sum_{m=0}^\infty \frac{\tildedeltaq_m}{\deltaq_m \deltaq_{m-1}} \ket{m}\bra{m} \nonumber\\
		&- \frac{\Omega^2}{4} \sum_{m=0}^\infty \sqrt{m+1}\sqrt{m+2} \Bigl(
		\frac{1}{\deltaq_{m}} - \frac{1}{\deltaq_{m+1}} \Bigr) \nonumber\\
		&\qquad\qquad\quad(e^{2i\phi}\ket{m}\bra{m+2}+\text{h.c.}),
	\end{align}
	where $\tildedeltaq_m= \deltaq -\anharm + \frac{g^2\Delta_{-1}\Delta_{3m}}{\Delta_{m}\Delta_{m-1}\Delta_{m-2}}$.
	Clearly the first term is the diagonal part while the second term is the off-diagonal one.
	With a similar procedure as the one used for~$S_1'$, it follows that
	\begin{align}
		S_2' = &\frac{\Omega^2}{8}e^{2i\phi} \sum_{m=0}^\infty \frac{\sqrt{m+1}\sqrt{m+2}}{\deltaq_{m}+\deltaq_{m+1}} \Bigl(
		\frac{1}{\deltaq_{m}} - \frac{1}{\deltaq_{m+1}} \Bigr) \nonumber\\
		&\qquad\qquad\quad\ket{m}\bra{m+2} - \text{h.c.}
		\label{eq:S2prime}
	\end{align}
	
	\begin{widetext}
		We can then compute~$S_3'$ from~\cref{eq:SW_S3_defeq}, i.e.
		\begin{align}
			\bbrkt{H_0^{\dressed},S_3'} &= \frac{1}{2}\bbrkt{S_2',H_{d1}^{\dressed}}_\offdiag + \frac{1}{3} \bbrkt{S_1', \bbrkt{S_1',H_{d1}^{\dressed}}_\diag}_\offdiag + \frac{1}{12} \bbrkt{S_1', \bbrkt{S_1',H_{d1}^{\dressed}}_\offdiag}_\offdiag.
		\end{align}
		The result is
		\begin{align}
			S_3' &= \Omega^3 e^{i\phi} \sum_{m=0}^\infty \ket{m}\bra{m+1} \Biggl( \frac{1}{12}\frac{\sqrt{m+1}}{(\deltaq_{m})^3} \Bigl( \frac{\tildedeltaq_{m+1}}{\deltaq_{m+1}} - \frac{\tildedeltaq_{m}}{\deltaq_{m-1}} \Bigr) \nonumber\\
			&\qquad\qquad\qquad\qquad\qquad\qquad+\frac{1}{96\deltaq_{m}}\Biggl(
			(m+2)\sqrt{m+1} \frac{\deltaq_{m}+4\deltaq_{m+1}}{\deltaq_{m+1}(\deltaq_{m}+\deltaq_{m+1})} \Bigl(
			\frac{1}{\deltaq_{m}} - \frac{1}{\deltaq_{m+1}} \Bigr) 
			\nonumber\\
			&\qquad\qquad\qquad\qquad\qquad\qquad\qquad\qquad- \sqrt{m+1}m \frac{4\deltaq_{m-1}+\deltaq_{m}}{\deltaq_{m-1}(\deltaq_{m-1}+\deltaq_{m})} \Bigl(
			\frac{1}{\deltaq_{m-1}} - \frac{1}{\deltaq_{m}} \Bigr) \Biggr)\Biggr) - \text{h.c.}\nonumber\\
			&\quad + \frac{\Omega^3}{96} e^{3i\phi} \sum_{m=0}^\infty \ket{m}\bra{m+3}  \frac{\sqrt{m+1}\sqrt{m+2}\sqrt{m+3}}{\deltaq_{m}+\deltaq_{m+1}+\deltaq_{m+2}} \nonumber \\
			&\quad\qquad\qquad\qquad    \Biggl(
			\frac{3\deltaq_{m+2}-\deltaq_{m+1}-\deltaq_{m}}{\deltaq_{m+2}(\deltaq_{m}+\deltaq_{m+1})} \Bigl(
			\frac{1}{\deltaq_{m}} - \frac{1}{\deltaq_{m+1}} \Bigr) 
			- \frac{3\deltaq_m-\deltaq_{m+1}-\deltaq_{m+2}}{\deltaq_m(\deltaq_{m+1}+\deltaq_{m+2})} \Bigl(
			\frac{1}{\deltaq_{m+1}} - \frac{1}{\deltaq_{m+2}} \Bigr)
			\Biggr)- \text{h.c.}
			\label{eq:S3prime}
		\end{align}
		
		We can eventually use~\cref{eq:S1prime,eq:S2prime,eq:S3prime} together with~\cref{eq:SWT_hamil_diag} to obtain~$H_0^{\ddressed}$ (defined in~\cref{eq:H_ddressed_full}):
		\begin{align}
			H_0^{\ddressed} &= \deltar \ar^\dagger\ar + \sum_{m=0}^\infty \ket{m}\bra{m}\Biggl( m\deltaq +\frac{\anharm}{2}m(m-1) + \frac{g^2m}{\Delta_{m-1}} \nonumber\\
			&\qquad\qquad\qquad\qquad\qquad\quad- \frac{\Omega^2\tildedeltaq_m}{4\deltaq_m \deltaq_{m-1}} 
			- \frac{\Omega^4}{32} \Biggl( \frac{m+1}{(\deltaq_{m})^3} \Bigl( \frac{\tildedeltaq_{m+1}}{\deltaq_{m+1}} - \frac{\tildedeltaq_{m}}{\deltaq_{m-1}} \Bigr) 
			- \frac{m}{(\deltaq_{m-1})^3} \Bigl( \frac{\tildedeltaq_{m}}{\deltaq_{m}} - \frac{\tildedeltaq_{m-1}}{\deltaq_{m-2}} \Bigr)
			\Biggr) \nonumber\\
			&\qquad\qquad\qquad\qquad\qquad\quad-\frac{\Omega^4}{192}\Biggl( \frac{1}{\deltaq_{m}} \Biggl(
			(m+2)(m+1) \frac{\deltaq_{m}+5\deltaq_{m+1}}{\deltaq_{m+1}(\deltaq_{m}+\deltaq_{m+1})} \Bigl(
			\frac{1}{\deltaq_{m}} - \frac{1}{\deltaq_{m+1}} \Bigr) 
			\nonumber\\
			&\qquad\qquad\qquad\qquad\qquad\qquad\qquad\qquad\quad- (m+1)m \frac{5\deltaq_{m-1}+\deltaq_{m}}{\deltaq_{m-1}(\deltaq_{m-1}+\deltaq_{m})} \Bigl(
			\frac{1}{\deltaq_{m-1}} - \frac{1}{\deltaq_{m}} \Bigr) \Biggr)\nonumber\\
			&\qquad\qquad\qquad\qquad\qquad\qquad\qquad-\frac{1}{\deltaq_{m-1}} \Biggl(
			(m+1)m \frac{\deltaq_{m-1}+5\deltaq_{m}}{\deltaq_{m}(\deltaq_{m-1}+\deltaq_{m})} \Bigl(
			\frac{1}{\deltaq_{m-1}} - \frac{1}{\deltaq_{m}} \Bigr) 
			\nonumber\\
			&\qquad\qquad\qquad\qquad\qquad\qquad\qquad\qquad\qquad- m(m-1) \frac{5\deltaq_{m-2}+\deltaq_{m-1}}{\deltaq_{m-2}(\deltaq_{m-2}+\deltaq_{m-1})} \Bigl(
			\frac{1}{\deltaq_{m-2}} - \frac{1}{\deltaq_{m-1}} \Bigr) \Biggr)\Biggr)\nonumber\\
			&\qquad\qquad\qquad\qquad\qquad\quad +\frac{\Omega^4}{96} \Biggl( 
			\frac{(m+2)(m+1)}{\deltaq_{m}+\deltaq_{m+1}} \Bigl(\frac{1}{\deltaq_{m}} - \frac{1}{\deltaq_{m+1}} \Bigr)^2
			- \frac{m(m-1)}{\deltaq_{m-2}+\deltaq_{m-1}} \Bigl(\frac{1}{\deltaq_{m-2}} - \frac{1}{\deltaq_{m-1}} \Bigr)^2
			\Biggr) \Biggr)  \nonumber\\
			&\quad -\ar^\dagger\ar \sum_m \frac{g^2\Delta_{-1}}{\Delta_m \Delta_{m-1}} \ket{m}\bra{m}.
			\label{eq:H0_ddressed}
		\end{align}
		\addition{We note that this expression implicitly contains all cross terms between the perturbative parameters~$g/\Delta$ and~$\Omega/\deltaq$ up to the chosen orders.}
		The approximate coupling Hamiltonian~$H_d^{\ddressed}$ (defined in~\cref{eq:H_ddressed_full}) up to 2nd~order in~$\Omega/\deltaq$ is instead given by
		\begin{align}
			H_d^{\ddressed} &= H_{d2}^{\dressed} + \bbrkt{S_1', H_{d2}^{\dressed}} + \bbrkt{S_2', H_{d2}^{\dressed}} + \frac{1}{2} \bbrkt{S_1',\bbrkt{S_1', H_{d2}^{\dressed}}} \\
			&\eqqcolon H_\mathrm{eff.coupl.}^{\ddressed} + H_\mathrm{resid.}^{\ddressed},
		\end{align}
		where
		\begin{align}
			H_\mathrm{eff.coupl.}^{\ddressed} &= e^{i\phi} a^\dagger \sum_{m=0}^\infty \ket{m}\bra{m+2} \Biggl(
			\tilde{g}_m \Bigl(1-\frac{\Omega^2}{8} \Bigl( \frac{m+3}{(\deltaq_{m+2})^2} + \frac{m+2}{(\deltaq_{m+1})^2} + \frac{m+1}{(\deltaq_{m})^2} + \frac{m}{(\deltaq_{m-1})^2} \Bigr) \Bigr) \nonumber\\
			&\qquad\qquad\qquad\qquad\qquad\quad + \frac{\Omega^2}{4} \Bigl(\frac{\sqrt{m+1}\sqrt{m+3}}{\deltaq_{m}\deltaq_{m+2}} \tilde{g}_{m+1}
			+ \frac{\sqrt{m}\sqrt{m+2}}{\deltaq_{m-1}\deltaq_{m+1}} \tilde{g}_{m-1} \Bigr) \nonumber\\
			&\qquad\qquad\qquad\qquad\qquad\quad+ \frac{\Omega^2}{4} \sqrt{m+1}\sqrt{m+2} \Biggl( \frac{g'_{m+2}}{\deltaq_{m}(\deltaq_{m}+\deltaq_{m+1})}
			- \frac{g'_{m+1}}{\deltaq_{m}\deltaq_{m+1}}
			+ \frac{g'_{m}}{\deltaq_{m+1}(\deltaq_{m}+\deltaq_{m+1})}
			\Biggr) \Biggr) + \text{h.c.} 
			\label{eq:H_effcoupl_ddressed}
		\end{align}
		with
		\begin{align}
			\tilde{g}_m &\coloneqq \frac{g\anharm\Omega \sqrt{m+1}\sqrt{m+2}}{2\Delta_{m}\Delta_{m+1}}\\
			g'_m &\coloneqq \frac{g\Omega\Delta_{-1}}{2\Delta_{m}\Delta_{m-1}},
		\end{align}
		and
		\begin{align}
			H_\mathrm{resid.}^{\ddressed} &= (e^{i\phi} \ar +\text{h.c.}) \sum_{m=0}^\infty \ket{m}\bra{m} \Biggl(
			g'_{m} \Bigl(1-\frac{\Omega^2}{4}   \Bigl( \frac{m+1}{(\deltaq_{m})^2} + \frac{m}{(\deltaq_{m-1})^2} \Bigr) \Bigr)  + \frac{\Omega^2}{4} \Bigl(\frac{m+1}{(\deltaq_{m})^2} g'_{m+1}
			+ \frac{m}{(\deltaq_{m-1})^2} g'_{m-1} \Bigr)
			\nonumber\\
			&\qquad\qquad\qquad\qquad\qquad\qquad\quad + \frac{\Omega^2}{4} \Bigl( \frac{\sqrt{m+1}\sqrt{m+2}\tilde{g}_m}{\deltaq_{m}(\deltaq_{m}+\deltaq_{m+1})}
			+ \frac{\sqrt{m}\sqrt{m+1}\tilde{g}_{m-1}}{\deltaq_{m}\deltaq_{m-1}}
			+ \frac{\sqrt{m-1}\sqrt{m}\tilde{g}_{m-2}}{\deltaq_{m-1}(\deltaq_{m-2}+\deltaq_{m-1})}
			\Bigr) \Biggr) \nonumber\\
			&\quad -\frac{\Omega}{2} e^{2i\phi} \ar \sum_{m=0}^\infty \ket{m}\bra{m+1} \frac{\sqrt{m+1}}{\deltaq_{m}}(g'_{m+1}-g'_{m}) +\text{h.c.}
			\nonumber\\
			&\quad -\frac{\Omega}{2} \ar^\dagger \sum_{m=0}^\infty \ket{m}\bra{m+1} \Bigl( \frac{\sqrt{m+1}}{\deltaq_{m}}(g'_{m+1}-g'_{m}) +\frac{\sqrt{m+2}}{\deltaq_{m+1}}\tilde{g}_{m} - \frac{\sqrt{m}}{\deltaq_{m-1}}\tilde{g}_{m-1} 
			\Bigr)  +\text{h.c.} \nonumber\\
			&\quad +\frac{\Omega^2}{4} e^{3i\phi} \ar \sum_{m=0}^\infty \ket{m}\bra{m+2} 
			\sqrt{m+1}\sqrt{m+2} \Biggl( \frac{g'_{m+2}}{\deltaq_{m}(\deltaq_{m}+\deltaq_{m+1})}
			- \frac{g'_{m+1}}{\deltaq_{m}\deltaq_{m+1}}
			+ \frac{g'_{m}}{\deltaq_{m+1}(\deltaq_{m}+\deltaq_{m+1})}
			\Biggr) +\text{h.c.} \nonumber\\
			&\quad -\frac{\Omega}{2} e^{2i\phi} \ar^\dagger \sum_{m=0}^\infty \ket{m}\bra{m+3} \Bigl( \frac{\sqrt{m+1}}{\deltaq_{m}}\tilde{g}_{m+1} - \frac{\sqrt{m+3}}{\deltaq_{m+2}}\tilde{g}_{m} 
			\Bigr)  +\text{h.c.} \nonumber\\
			&\quad +\frac{\Omega^2}{4} e^{3i\phi} \ar^\dagger \sum_{m=0}^\infty \ket{m}\bra{m+4} 
			\Biggl( \frac{\sqrt{m+1}\sqrt{m+2}\tilde{g}_{m+2}}{\deltaq_{m}(\deltaq_{m}+\deltaq_{m+1})}
			- \frac{\sqrt{m+4}\sqrt{m+1}\tilde{g}_{m+1}}{\deltaq_{m}\deltaq_{m+3}}
			+ \frac{\sqrt{m+3}\sqrt{m+4} \tilde{g}_{m}}{\deltaq_{m+3}(\deltaq_{m+3}+\deltaq_{m+2})}
			\Biggr)+\text{h.c.}
			\label{eq:H_resid_ddressed}
		\end{align}
	\end{widetext}
	All terms in~$H_\mathrm{resid.}^{\ddressed}$ are relatively small and off-resonant with the~$\ket{20}\leftrightarrow\ket{01}$ transition so we expect them to have a small effect and we do not proceed with higher orders of ~\SW s.
	
	\subsection{Analysis of the $\ket{20}\leftrightarrow\ket{01}$ avoided crossing}
	\label{sub:starkshift_effcouplig}
	
	In this section we give the methods used to calculate the curves in~\cref{fig1:readout_res_analytical}(c),(e).
	
	We define~$\omega_d^*$ as the drive frequency corresponding to the center of the~$\ket{20}\leftrightarrow\ket{01}$ avoided crossing of the full Hamiltonian~$H$ as given in~\cref{eq:H_full_main}.
	Then the exact value of the effective $\ket{20}\leftrightarrow\ket{01}$ coupling~$\tilde{g}$ is given by half the energy separation at that point.
	The avoided crossing can be found numerically by exact diagonalization as a function of~$\omega_d$.
	
	In the subspace~$\drivesub=\linspan\{\ket{20},\ket{01}\}$ we can write~$H$ as $H|_{\drivesub}\equiv-\eta(\omega_d)\ztype/2+\tilde{g}(\omega_d)[\cos(\phi)\xtype+\sin(\phi) \ytype]=-\eta(\omega_d)\ztype/2+\tilde{g}(\omega_d)\xtype$ for~$\phi=0$ as in~\cref{sub:analytical_results_main}.
	As we want to implement a $\ket{20}\leftrightarrow\ket{01}$ $\pi$~rotation, we notice that the choice of~$\phi$, i.e.~the choice of rotation axis in the equator of the Bloch sphere, is irrelevant.
	We have also ignored a term proportional to the identity~$\idtype$, which gives a phase difference with respect to states outside of~$\drivesub$, in particular between the computational and leakage subspaces of the transmon.
	However, this phase is irrelevant if~$\ket{20}$ is swapped entirely onto~$\ket{01}$ since the latter decays and dephases fast, thus suppressing any phase coherence.
	As demonstrated in~\cref{sub:sims_readres} the res-LRU can reach a very high~$\leakredrate$, for which the effect of this phase is then minimal.
	Assuming that~$H_\mathrm{resid.}^{\ddressed}$ in~\cref{eq:H_resid_ddressed} is negligible, an analytical approximation of~$\eta$ is given by
	\begin{align}
		\eta(\omega_d) \approx \braket{20|H_0^{\ddressed}(\omega_d)|20} - \braket{01|H_0^{\ddressed}(\omega_d)|01},
		\label{eq:eta_suppl}
	\end{align}
	where we have made the dependence of~$H_0^{\ddressed}$ in~\cref{eq:H0_ddressed} on~$\omega_d$ explicit.
	This holds since then~$H_0^{\ddressed}$ accounts for all the Stark shifts of~$\ket{20}$ and~$\ket{01}$ due to the capacitive coupling and the drive (up to the given orders).
	The center of the avoided crossing is found by imposing the condition~$\eta(\omega_d)=0$.
	As the explicit expression that can be extracted from~\cref{eq:H0_ddressed} is not analytically solvable, we use the secant method available in~\emph{scipy} to find~$\omega_d^*$ that fulfills this condition in~\cref{eq:eta_suppl}.
	It is then straightforward to compute the (approximate) analytical estimate for the effective coupling as $\tilde{g}(\omega_d^{*})=|\braket{01|H_\mathrm{eff.coupl.}^{\ddressed}(\omega_d^{*})|20}|$ from~\cref{eq:H_effcoupl_ddressed}, which is plotted in~\cref{fig1:readout_res_analytical}(e).
	
	\section{Further characterization of the readout-resonator LRU}
	\label{sec:details_FT_sims}
	
	\subsection{Effective $\Tone$ and~$\Ttwo$ due to the drive}
	\label{sub:more_heatmaps}
	
	\begin{figure}
		\centering
		\includegraphics{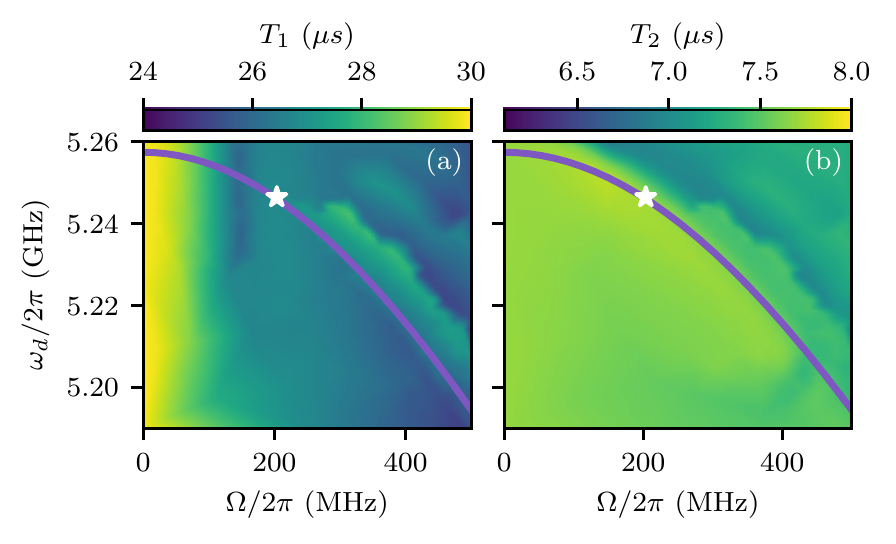}
		\caption{\label{fig:T1_T2}
			Effective~$\Tone$~(a) and~$\Ttwo$~(b) which account for the extra decoherence caused by the drive during the time slot~$\Tslot=440~\ns$.
			We can see that the variation is small as a function of the drive amplitude compared to the values at~$\Omega=0$.
			The white star indicates the chosen operating point \addition{($\Omega/2\pi\approx204~\MHz$, $\omega_d/2\pi\approx5.2464~\GHz$, $\tpulse=178.6~\ns$, see~\cref{sub:sims_readres})}.
			The purple line corresponds to the higher order estimate of the optimal drive frequency~$\omega_d^*$ as a function of~$\Omega$ (see~\cref{fig1:readout_res_analytical}(c)).
			The heatmaps are sampled using the \emph{adaptive} package~\cite{Nijholt19}.
		}
	\end{figure}
	
	In this section we discuss the effects of the readout-resonator LRU within the computational subspace when applied to a non-leaked transmon.
	As pulses at different $(\omega_d,\Omega)$~points have a different duration~$\tpulse$, it would not be fair to report an effective~$\Tone$ and~$\Ttwo$ during~$\tpulse$.
	That is, stronger pulses potentially produce lower~$\Tone$ and~$\Ttwo$, but they also take less time to implement the LRU.
	However, the overall disturbance to the qubit is a combination of these two factors.
	We thus report an effective~$\Tone$ and~$\Ttwo$ during the whole time slot of~$\Tslot=440~\ns$, leading to a uniform metric for the whole $(\omega_d,\Omega)$~landscape.
	Specifically, to estimate~$\Tone$ we prepare the state~$\ket{1}\bra{1}\otimes\thermal$, we simulate the Lindblad equation in~\cref{eq:Lindblad} and we evaluate the remaining population~$\onepop$ in~$\ket{1}$ at the end of the time slot after tracing out the resonator.
	Assuming that $\onepop=e^{-\Tslot/\Tone}$ we then compute~$\Tone$ by inverting this formula.
	To estimate~$\Ttwo$ we prepare $\ket{+}\bra{+}\otimes\thermal$ and we evaluate the decay of the off-diagonal transmon matrix element~$\ket{0}\bra{1}$ as this is directly available in simulation (rather than simulating a full Ramsey experiment).
	We then invert $\abs{\braket{0|\tr_r(\rho(\Tslot))|1}}=e^{-\Tslot/\Ttwo}/2$ to get~$\Ttwo$.
	
	\Cref{fig:T1_T2}~shows the resulting effective~$\Tone$ and~$\Ttwo$.
	\addition{In~\cref{fig:T1_T2}(a) one can see that~$\Tone$ decreases by at most~15\% as a function of~$\Omega$, showing that a short~$\tpulse$ mostly counterbalances the effect of a strong~$\Omega$.
	In particular, $\Tone\approx27.1~\us$ at the operating point.
	On the other hand, one can notice that~$\Tone$ dips around~$\Omega_\text{cr}/2\pi=143~\MHz$, where the pulses are very long, suggesting that driving slightly into the underdamped regime is favourable.}
	In~\cref{fig:T1_T2}(b) one can see that the value of~$\Ttwo$ is about~\addition{$7.7~\us$} at~$\Omega=0$, i.e.~when no pulse is applied.
	This has to be contrasted with the input $\Ttwo$~parameter of~$30~\us$ inserted in the Lindblad equation (see~\cref{tab:target_parameters}).
	We assume that that implicitly accounts for dephasing caused by flux noise only.
	Photon-shot noise from the resonator is a further dephasing source which is explicitly included in these simulations.
	The combination of flux and photon-shot noise leads to the actual effective~$\Ttwo$ reported in~\cref{fig:T1_T2}(b).
	We note that if~$\avgphoton=0$ then the effective~$\Ttwo$ at~$\Omega=0$ would exactly match the input of~$30~\us$.
	While the effective~$\Ttwo$ can be restored from~\addition{$7.7~\us$} to~$30~\us$ with colder resonators or by engineering different system parameters altogether, the important information from~\cref{fig:T1_T2}(b) is that~$\Ttwo$ barely changes as a function of~$\Omega$.
	Combined with the similar result for~$\Tone$, this means that the drive causes only a marginal effect within the computational subspace.
	Notice that in the region where the readout-resonator LRU is most effective (just above the purple line in~\cref{fig:T1_T2}(b)), $\Ttwo$~is even slightly higher than at~$\Omega=0$ (\addition{$7.9$}~versus~\addition{$7.7~\us$}).
	We attribute this to the fact that the pulse temporarily reduces the excited-state population in the resonator (see~\cref{fig2:readout_res_sims}(d)).
	In this way photon-shot noise is reduced until the resonator re-thermalizes, however at the cost of some leakage of the transmon.
	
	\addition{In~\cref{fig2:readout_res_sims}(d) one can notice that a non-negligible amount of population ends up in~$\ket{10}$ from the initial state~$\ket{0}\bra{0}\otimes\thermal$.
	This corresponds to an excitation rate~$\Toneup\approx256~\us$ at the operating point.
	We backtrack this source of error to a combination of the drive and the jump operator~$a^\dagger$, corresponding to the drive inducing a transmon excitation rate based on the resonator excitation rate.
	However, as here $\Toneup\gg\max\{\Tone,\Ttwo\}$, it is not a limiting factor and we have not included it in the Surface-17 simulations.
	}
	
	\subsection{Long-drive limit \addition{in the underdamped regime} and its drawbacks as a LRU}
	\label{sub:eth_approach}
	
	\begin{figure}
		\centering
		\includegraphics{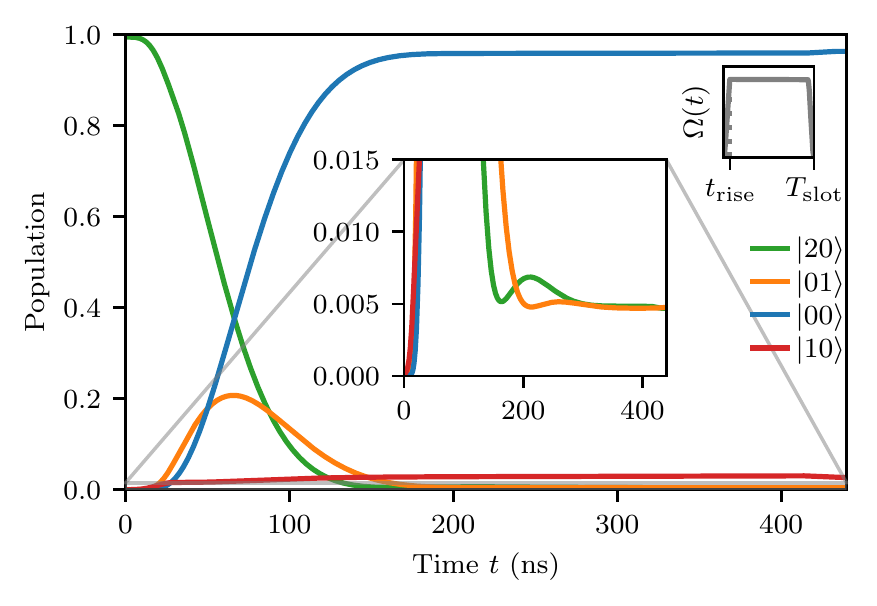}
		\caption{\label{fig:long_drive}
			Time evolution from the initial state~$\ket{2}\bra{2}\otimes\thermal$ for~$\trise=30~\ns$ and for an otherwise always-on drive during~$\Tslot$.
			This is simulated with the same~$\Omega/2\pi\approx204~\MHz$ and~$\omega_d/2\pi\approx5.2464~\GHz$ as at the operating point in~\cref{fig2:readout_res_sims}.
		}
	\end{figure}
	
	In this section we compare the reset schemes in~\cite{Zeytinoglu15,Egger18} versus~\cite{Magnard18} in terms of their performance as a LRU \addition{in the underdamped regime}.
	The approach of~\cite{Zeytinoglu15,Egger18}, which we have adopted in~\cref{sub:sims_readres}, aims at swapping~$\ket{20}$ and~$\ket{01}$ by targeting the first minimum of the oscillations induced by the drive (switching the drive off afterwards).
	As shown in~\cref{sub:sims_readres}, this approach allows for a residual leakage population~\addition{$\leakpop_\mathrm{op.}\approx0.5\%$} at the operating point (see~\cref{fig2:readout_res_sims}(a)), given our parameters (see~\cref{tab:target_parameters}).
	\addition{While this already reaches thermal-state levels (here~$\avgphoton=0.5\%$) with the considered system parameters,} the approach in~\cite{Magnard18} could be used in general to achieve an even lower or similar~$\leakpop$ \addition{(in particular for lower~$\kappa$'s)}.
	
	The approach in~\cite{Magnard18} keeps the drive on for a much longer period of time \addition{(at least one more oscillation)} allowing both the populations in~$\ket{20}$ and~$\ket{01}$ to decay to almost~0, modulo thermal excitations.
	\Cref{fig:long_drive}~shows that it is indeed possible to suppress these populations to thermal-state levels, where we use the same $(\Omega,\omega_d)$ as at the operating point (see~\cref{sub:sims_readres}).
	However, we see that for the operating point there is almost no gain by using this approach.
	Furthermore, this approach costs much more time \addition{and could exceed~$\Tslot=440~\ns$ if~$\kappa$ is not as high as assumed here.
	In particular, in that case the first few minima after the first one could be slightly higher, due to transmon decoherence, and one would need to wait even longer to overcome this effect.}
	
	Another disadvantage of the approach in~\cite{Magnard18} is that the disturbance to the qubit is stronger as the drive is kept on for a longer period of time.
	E.g.,~in~\cref{fig:long_drive} one can see that~$\ket{00}$ and~$\ket{10}$ reach an equilibrium thanks to the drive (even in the presence of relaxation), where the population in~$\ket{10}$ is \addition{higher than in~\cref{fig2:readout_res_sims}(b).
	By evaluating~$\Tone$ we find $\Tone\approx23~\us$ instead of~$27~\us$ (see~\cref{sub:more_heatmaps}).}
	Furthermore, \addition{if one would have to use a $\tpulse>\Tslot$ when~$\kappa$ is lower than here}, then the QEC~cycle would get longer, affecting the coherence of all qubits, not only of the high-frequency data qubits to which the \resLRU~is applied.
	
	\addition{
	\subsection{Sensitivity to residual $ZZ$~crosstalk}
	\label{sub:residualZZ}
	
	\begin{figure}
		\centering
		\includegraphics{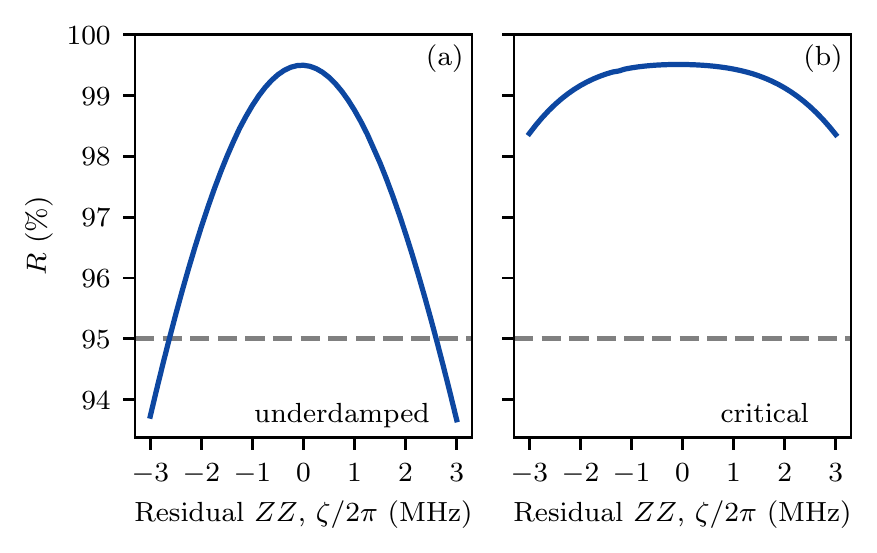}
		\caption{\label{fig:residualZZ}
			\addition{
			Sensitivity of the leakage-reduction rate~$\leakredrate$ of the readout-resonator LRU as a function of the overall residual $ZZ$~coupling~$\zeta$.
			(a)~Underdamped regime, specifically at the operating point ($\Omega/2\pi\approx204~\MHz$, $\omega_d/2\pi\approx5.2464~\GHz$, $\tpulse=178.6~\ns$, see~\cref{sub:sims_readres}).
			(b)~Critical regime ($\Omega/2\pi\approx143~\MHz$, $\omega_d/2\pi\approx5.252~\GHz$, $\tpulse=440~\ns$).
			}
		}
	\end{figure}

	In a multi-transmon chip, each transmon is coupled to one or more neighbors.
	In general, if the coupling is not tunable there can be some residual $ZZ$~crosstalk, i.e.~a shift of the transmon frequency by an amount~$\zeta$ based on whether each neighboring transmon is in~$\ket{1}$ instead of~$\ket{0}$.
	In this section we study the effect of this $ZZ$~coupling on the readout-resonator LRU, which we assume being tuned up when all neighbors are in~$\ket{0}$.
	We do not include neighboring transmons in our simulations, so we mimic it by shifting the transmon frequency (while keeping the drive parameters fixed).
	
	In~\cref{fig:residualZZ} we perform the analysis for the operating point (see~\cref{sub:sims_readres}), which resides in the underdamped regime, and for the critical point.
	In both cases the leakage-reduction rate~$\leakredrate$ scales seemingly quadratically.
	In the underdamped regime the pulse targets the first minimum of the damped Rabi oscillations, so it is more sensitive to a variation in frequency than in the critical regime.
	However, we can observe that for~$\abs{\zeta}/2\pi\lesssim2~\MHz$ (note that this is the cumulative $ZZ$~coupling over all neighbors) $\leakredrate$~stays above~95\%, which is the conservative value we have used in~\cref{sub:logical_performance} and for which the logical error rate was already close to optimal in Surface-17 (see~\cref{sub:logerrrate_vs_LRUpars}).
	Regarding other performance parameters of the LRU, we find that~$\leakrateLRU$ scales in the same relative way as~$\leakredrate$ by unitarity, whereas~$\Tone,\Ttwo$ and~$\Toneup$ vary by~$\lesssim1\%$.
	}

	\section{Further Surface-17 characterization}
	\label{sec:s17_suppl}
	
	\subsection{Details about the density-matrix simulations}
	\label{sub:sim_params}
	
	The parameters used in this work are reported in~\cref{tab:sim_params}.
	
	\begin{table}
		\begin{tabular}{@{} *5l @{}}    \toprule
			Parameter                                          & Value       & \\ \midrule
			\hline \hline
			Relaxation time \(\Tone\)                          & 30~\(\us\)    \\
			\hline
			Sweetspot pure-dephasing time~\(\tdephsweet\)           & 60~\(\us\)    \\
			\hline
			High-freq.~pure-dephasing time                                          \\ at interaction point~\(\tdephint\) & 8~\(\us\) \\
			\hline
			Mid-freq.~pure-dephasing time                                           \\ at interaction point~\(\tdephint\) & 6~\(\us\) \\
			\hline
			Mid-freq.~pure-dephasing time                                           \\ at parking point~\(\tdephpark\) & 8~\(\us\) \\
			\hline
			Low-freq.~pure-dephasing time                                           \\ at parking point~\(\tdephpark\) & 9~\(\us\) \\
			\hline
			Single-qubit gate time~\(\tgate\)                  & 20~\(\ns\)    \\
			\hline
			Two-qubit interaction time~\(\tint\)               & 30~\(\ns\)    \\
			\hline
			Single-qubit phase-correction time~\(\tphasecorr\) & 10~\(\ns\)    \\
			\hline
			Readout-resonator LRU time~\(\tresLRU\) & 100~\(\ns\)    \\
			\hline
			$\ket{1}\leftrightarrow\ket{2}$ $\pi$-pulse time~\(\tpiLRU\) & 20~\(\ns\)    \\
			\hline
			Measurement time~\(\tmeas\)                        & 580~\(\ns\)   \\
			\hline
			QEC-cycle time~\(\tcycle\)                         & 800~\(\ns\)   \\ \bottomrule
			\hline
		\end{tabular}
		\caption{\label{tab:sim_params}
			The parameters for the qubit coherence times and for the gate, LRU, measurement and \QEC-cycle durations used in the density-matrix simulations.
			The interaction point corresponds to the frequency to which a transmon is fluxed to implement a~$\CZ$, whereas the parking point to the frequency at which the ancilla qubits are parked during measurement~\cite{Versluis17}.
		}
	\end{table}

	\subsubsection{\resLRU~in \emph{quantumsim}}
	\label{subsub:resLRU_quantumsim}
	
	A comprehensive review of the density-matrix simulations and the use of the \emph{quantumsim} package~\cite{quantumsim_website} is available at~\cite{Obrien17,Varbanov20}.
	In this section we explain the specific implementation of the newly introduced~\resLRU, expressed in the Pauli Transfer Matrix formalism.
	
	\begin{figure*}
		\centering
		\includegraphics{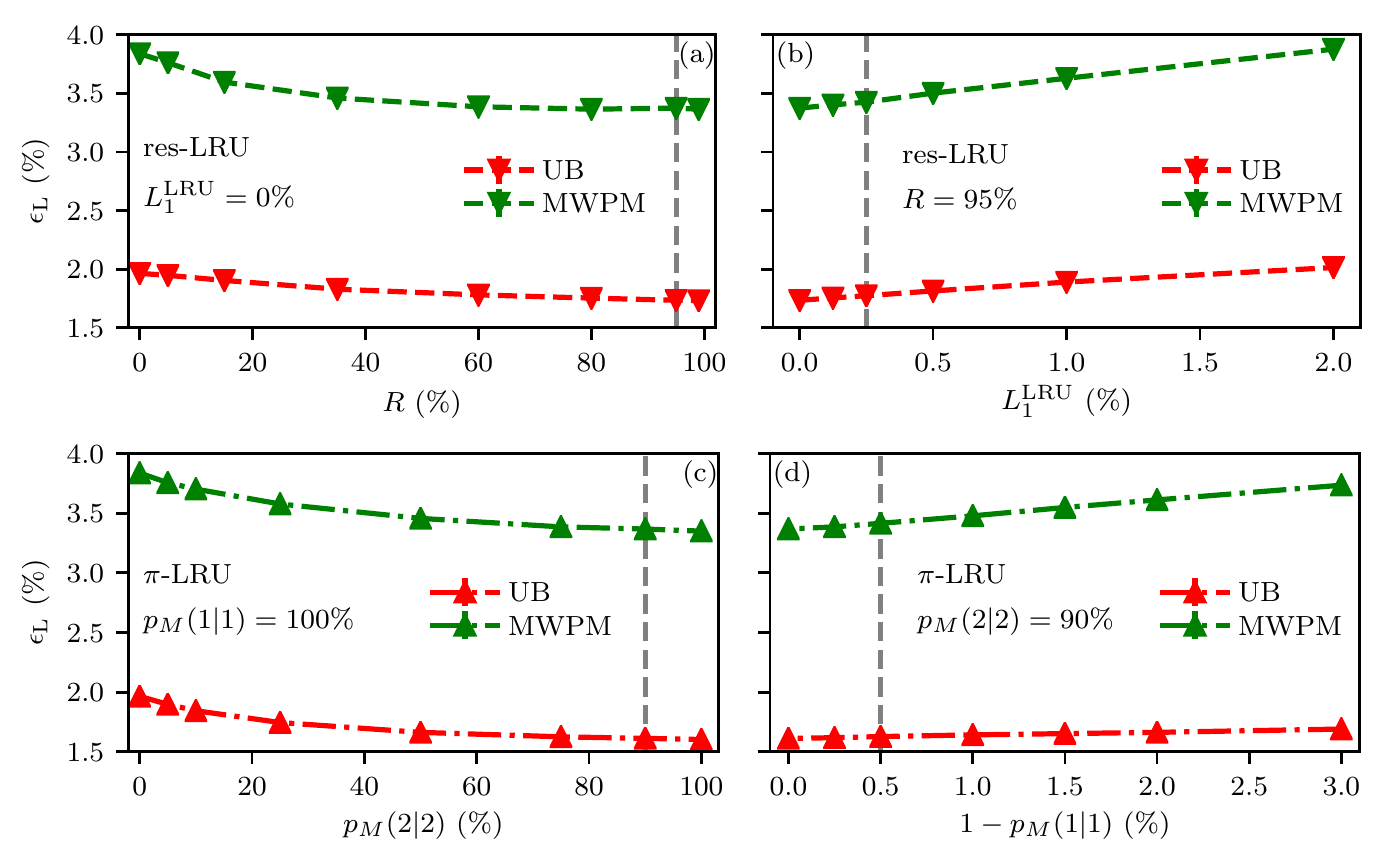}
		\caption{\label{fig:further_logerrrate}
			Logical error rate~$\logerrrate$ per QEC~cycle as a function of various LRU~parameters.
			(a),(b)~use only the~\resLRU, while~(c),(d) the \piLRU.
			We fix~$\leakrate=0.5\%$ for all.
			Vertical dashed lines indicate the values considered in~\cref{sub:logical_performance}.
			These results are extracted from $2\times10^4$~runs of 20~QEC~cycles each per choice of parameters.
			Error bars are estimated using bootstrapping and are smaller than the symbol size.
		}
	\end{figure*}
	
	We construct a \quotes{phenomenological} Lindblad model with input parameters~$\leakredrate,\leakrateLRU$ and~$\tresLRU$.
	We use the Pauli Transfer Matrix~$S_\text{\resLRU}=S_{\uparrow} S_{\downarrow}$, where~$S_{\downarrow}$ is the Pauli Transfer Matrix of the superoperator~$\mathcal{S}_{\downarrow}=e^{\tresLRU \mathcal{L}_{\downarrow}}$ and the Lindbladian~$\mathcal{L}_{\downarrow}$ has the quantum jump operator
	\begin{align}
		K_{\downarrow}=\frac{1}{\sqrt{\frac{\tresLRU}{-\log(1-\leakredrate_\mathrm{sim})}}}\ket{0}\bra{2}
	\end{align}
	with~$\leakredrate_\mathrm{sim}$ to be determined.
	Besides this, $\mathcal{L}_{\downarrow}$ has the standard qutrit jump operators for relaxation and dephasing~\cite{Varbanov20}.
	On the other hand, $S_{\uparrow}$ is the Pauli Transfer Matrix of the superoperator~$\mathcal{S}_{\uparrow}=e^{\mathcal{L}_{\uparrow}}$ and the Lindbladian~$\mathcal{L}_{\uparrow}$ has a single jump operator
	\begin{align}
		K_{\uparrow}=\frac{1}{\sqrt{\frac{1}{-\log(1-2\leakrateLRU)}}}\ket{2}\bra{0}
	\end{align}
	since relaxation and dephasing during~$\tresLRU$ are already accounted for by~$S_{\downarrow}$.
	In this way, calling~$p^{\ket{j}}_{i}, p^{\ket{j}}_{f}$ the populations before and after the \resLRU, if we apply~$S_\text{\resLRU}$ on a non-leaked transmon we get~$\leakpop_{f}=2\leakrateLRU \zeropop_{i}$, consistently with~\cref{subsub:resLRU_model_qsim}.
	Instead, if we apply~$S_\text{\resLRU}$ to a leaked transmon~($\leakpop_{i}=1$) we get $\leakpop_{f}\approx 1-\leakredrate_\text{sim}+2\leakrateLRU$.
	By fixing~$\leakredrate_\text{sim}=\leakredrate+2\leakrateLRU$ we match the definition of~$\leakredrate$ in~\cref{subsub:resLRU_model_qsim} as well.
	The approximation is very good for large~$\leakredrate$ and low~$\leakrateLRU$, which is precisely the interesting regime for \resLRU~that we have explored.
	
	\addition{\subsubsection{Decoding}
	\label{subsub:decoding}
	
	In this section we provide additional information on the \UB~and \MWPM~decoders~\cite{Obrien17,Obrien19}.
	
	UB considers the 32~computational states that differ by a purely $\xtype$~error on top of~$\ket{0}_\mathrm{L}$ and that are independent (i.e.~they cannot be obtained from each other by multiplication with an $\xtype$-type stabilizer).
	At the end of each QEC~cycle~$\cycle$, each possible final $\ztype$~syndrome is compatible with a pair of these states, where one can be associated with~$\ket{0}_\mathrm{L}$ and the other with~$\ket{1}_\mathrm{L}$ as they differ by the application of any representation of~$\xtype_\mathrm{L}$.
	The largest overlap of these two states with the diagonal of the density matrix at QEC~cycle~$\cycle$ corresponds to the maximum probability of correctly guessing whether a $\xtype_\mathrm{L}$~error has occurred or not upon performing a logical measurement of~$\ztype_\mathrm{L}$.
	The latter is assumed to be performed by measuring all data qubits in the $\{\ket{0},\ket{1},\ket{2}\}$~basis and computing the overall parity.
	To compute the parity we assume that a~$\ket{2}$ is declared as a~$\ket{1}$ since decoders usually do not use information about leakage (and since measurements often declare~$\ket{2}$ as a~$\ket{1}$ rather than as a~$\ket{0}$).
	Then UB~computes~$\logicalfid(\cycle)$ by weighing this probability with the chance of measuring the given final $\ztype$~syndrome (conditioned on the density matrix) and by summing over all possible syndromes.
	In other words, UB~always finds the correction that maximizes the likelihood of the logical measurement returning the initial state, here~$\ket{0}_\mathrm{L}$.
	As UB~uses information generally hidden in the density matrix, it gives an upper bound to the performance of any realistic decoder, which can at most use the syndrome information extracted via the ancilla qubits.
	
	MWPM tries to approximate the most likely correction by finding the lowest weight correction, which is a good approximation when physical error rates are relatively low.
	As the ancilla qubits can be faulty, the decoding graph is three dimensional.
	In particular, we allow for space-like edges corresponding to data-qubit errors, time-like edges corresponding to ancilla-qubit errors and spacetime-like edges corresponding to data-qubit errors occurring in the middle of the parity-check circuit.
	The weights are extracted with the adaptive algorithm in~\cite{Spitz18} from a simulation ($10^5$~runs of 20~QEC~cycles each) without leakage and an otherwise identical error model.
	Similarly to \UB, for decoding we assume that a~$\ket{2}$ is declared as a~$\ket{1}$ since the standard \MWPM~does not account for leakage.
	}
	
	\subsection{Logical error rate as a function of the LRU parameters}
	\label{sub:logerrrate_vs_LRUpars}
	
	We study the variation in the logical error rate~$\logerrrate$ per QEC~cycle as a function of the performance parameters of the LRUs.
	Here we fix~$\leakrate=0.5\%$ as it is easier to visualize variations in~$\logerrrate$ with a relatively large~$\leakrate$.
	The leakage-reduction rate~$\leakredrate$ and the readout probability~$\readoutp{2}{2}$ play similar roles for the \resLRU~and \piLRU, respectively.
	In~\cref{fig:further_logerrrate}(a),(c) one can see that this is the case and that the values of~$\logerrrate$ at the parameters used in~\cref{sub:logical_performance} ($R=95\%$ and~$\readoutp{2}{2}=90\%$) are very close to their best values (at least for this system size).
	This shows that the advantages of a larger~$R$ or~$\readoutp{2}{2}$ are marginal.
	We attribute this to the fact that leakage is exponentially suppressed with an already quite large exponent.
	Furthermore, the parameters~$\leakrateLRU$ and~$1-\readoutp{1}{1}=\readoutp{2}{1}$, regulating the induced leakage, play similar roles as well, as~\cref{fig:further_logerrrate}(b),(d) show.
	We see that~$\logerrrate$ is more sensitive to~$\leakrateLRU$ and~$1-\readoutp{1}{1}$ compared to~$\leakredrate$ and~$\readoutp{2}{2}$.
	In particular we see that~$\logerrrate$ is slightly larger at the parameters used in~\cref{sub:logical_performance} ($\leakrateLRU=0.25\%$ and~$1-\readoutp{1}{1}=0.5\%$) rather than at~0, although the difference is small.
	
	\addition{
	\subsection{Effect of the leakage conditional phases on the logical error rate}
	\label{sub:leakcondphases}
	
	\begin{figure*}
		\centering
		\includegraphics{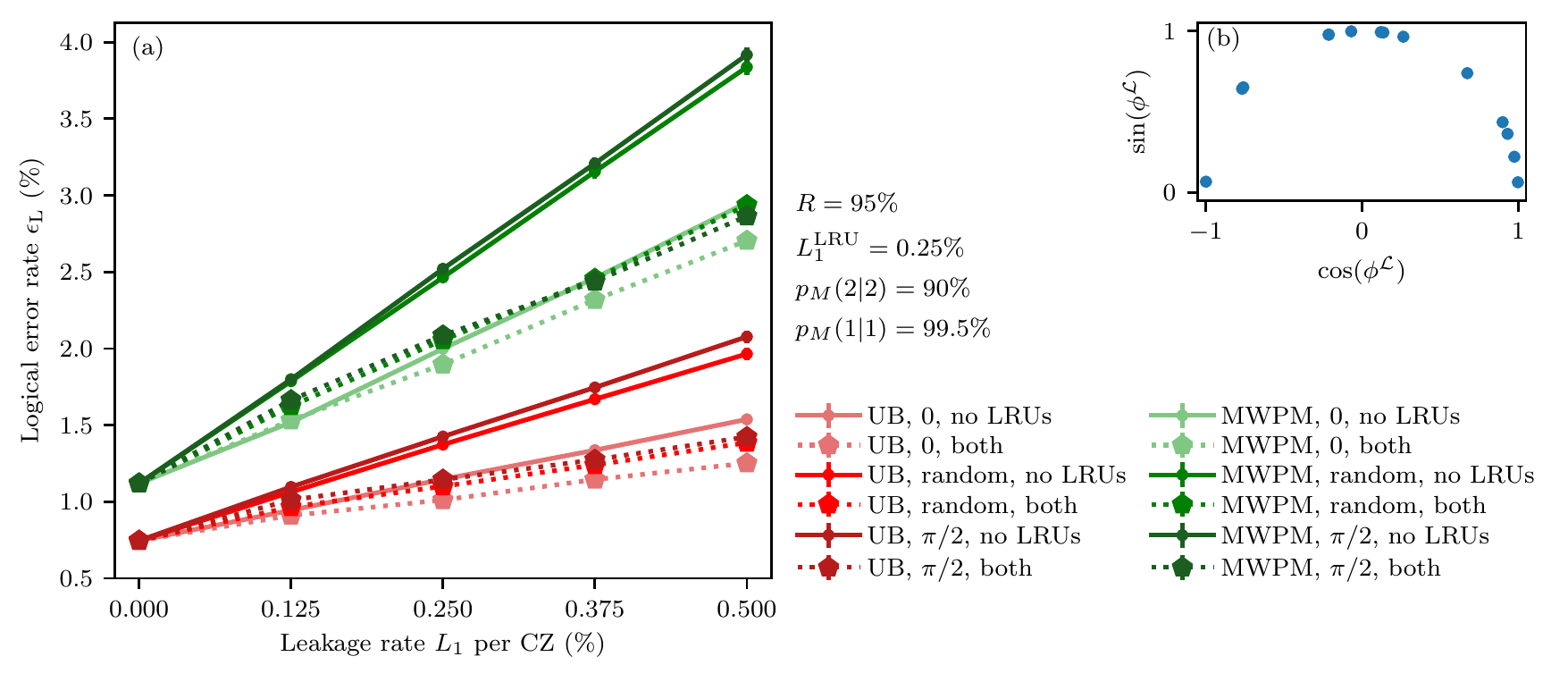}
		\caption{\label{fig:leakcondphases}\addition{
			Variation of the logical error rate~$\logerrrate$ for different choices of leakage conditional phases~$\leakcondphase$.
			(a)~$\logerrrate$ per QEC~cycle for~\UB~(shades of red) and~\MWPM~(shades of green) versus~$\leakrate$, in the cases with: no LRUs and both LRUs, each for all~$\leakcondphase$ set to 0,~$\pi/2$~or uniformly random in~$[0,\pi]$.
			These results are extracted from $2\times10^4$~runs of 20~QEC~cycles each per choice of parameters.
			Error bars are estimated using bootstrapping and are mostly smaller than the symbol size.
			(b)~The random values for~$\leakcondphase$ used across this work.
			These values are extracted from a uniform distribution in~$[0,\pi]$.
			We have excluded negative values as~$\pm\leakcondphase$ corresponds to the same chance of spreading a $\ztype$~error under the twirling action of the parity-check measurements.
		}}
	\end{figure*}
	
	As defined in the main text the leakage conditional phases are the phases that a non-leaked transmon acquires when interacting with a leaked one during a~$\CZ$.
	Here we denote them as~$\leakcondphase[\flux]$ and~$\leakcondphase[\static]$ depending on whether the lower or the higher frequency transmon of the pair is leaked, respectively, and we use~$\leakcondphase$ to indicate either of them.
	Furthermore, in this section we use the notation $\ket{\text{low-f.~transmon}, \text{high-f.~transmon}}$.
	Note that for a~$\CZ$ between two qutrits in principle there are~9 phases ($\phi_{00},\phi_{01},\phi_{10},\phi_{11},\phi_{02},\phi_{20},\phi_{21},\phi_{12},\phi_{22}$), where the first~4 are fixed to $0,0,0,\pi$, respectively.
	Of the 5~phases containing a~$\ket{2}$ we consider only two of them here, i.e.~$\leakcondphase[\static]=\phi_{02}-\phi_{12}$ and~$\leakcondphase[\flux]=\phi_{20}-\phi_{21}$ as defined above.
	This is because in our leakage model~\cite{Varbanov20} we set to~0 the coherence between the computational and leakage subspace of each qutrit, motivated by the fact that leakage is projected relatively fast and that the stabilizer measurements ideally prevent any interference effect.
	This means that the individual phases are global phases, whereas their difference cannot be gauged away when the non-leaked qubit is in a superposition of~$\ket{0}$ and~$\ket{1}$.
	
	For a flux-based~$\CZ$ with conditional phase~$\pi$ for~$\ket{11}$, ideally one should have~$\leakcondphase[\flux]=0$ and~$\leakcondphase[\static]=\pi$~\cite{Varbanov20} as~$\ket{02}$ acquires a conditional phase equal and opposite to~$\ket{11}$.
	If only~$\ket{12}$ and~$\ket{21}$ are coupled in the 3-excitation manifold, it holds~$\leakcondphase[\static]=\pi-\leakcondphase[\flux]$.
	The strength of the repulsion times the $\CZ$~duration gives e.g.~$\leakcondphase[\flux]\sim\pi/4$ for the parameters in~\cite{Varbanov20}.
	However, $\ket{03}$~interacts with~$\ket{12}$ and~$\ket{21}$ and breaks the relationship above, for which we can consider~$\leakcondphase[\flux]$ and~$\leakcondphase[\static]$ as effectively unconstrained.
	The randomized values used across the main text are reported in~\cref{fig:leakcondphases}(b).
	We use 14~values, of which 3 for~$\leakcondphase[\static]$ and 3~for~$\leakcondphase[\flux]$ when each high-frequency data qubit is leaked or interacts with a leaked ancilla qubit, respectively, and 8 only for~$\leakcondphase[\static]$ when each ancilla qubit is leaked and interacts with a low-frequency data qubit (as low-frequency data qubits cannot leak themselves).
	
	In this section we study the dependence of the logical error rate~$\logerrrate$ on the leakage conditional phases, without discussing how one would engineer the system to tune them to certain values.
	The best-case scenario to minimize~$\logerrrate$ is to set all~$\leakcondphase=0$, since no $\ztype$~rotations are spread then.
	Instead, the worst-case scenario corresponds to all~$\leakcondphase=\pi/2$, since under the twirling effect of the parity-check measurements this corresponds to spreading a $\ztype$~error with $50\%$~chance.
	Notice that, if all~$\leakcondphase=\pi$, overall the spread errors amount to a stabilizer (except in the QEC~cycle in which leakage occurs), so it is close to the best-case scenario.
	
	\Cref{fig:leakcondphases}(a)~compares the logical performance for both~UB and~MWPM in the cases where $\leakcondphase=0$, $\leakcondphase=\pi/2$ and when they are random as in~\cref{fig5:logical_performance} and in the rest of this work.
	First, one can notice that the performance of random~$\leakcondphase$ is very close to the worst-case scenario~($\leakcondphase=\pi/2$).
	This is due to the fact that it is not necessary to spread an error on every qubit with $50\%$~chance each to cause a logical error with high probability.
	Second, one can see that just tuning all~$\leakcondphase=0$ without implementing~LRUs is almost as good (or even better) as using the LRUs when~$\leakcondphase$ are random.
	We attribute this to the fact that one of the major effects of the LRUs is to prevent correlated errors being spread by a leaked qubit for many QEC~cycles.
	Tuning~$\leakcondphase=0$ achieves this as well, but it still does not address the fact that the code distance is effectively reduced if a data qubit stays leaked and that the full stabilizer information is not accessible as long as an ancilla qubit is leaked.
	Indeed, using LRUs even when~$\leakcondphase=0$ always allows for a lower logical error rate (see~\cref{fig:leakcondphases}(a)).
	Furthermore, the reduction in distance and the corruption of the stabilizer information suggest that a threshold would still likely be low without using LRUs.
	}
	

%

\end{document}